\documentstyle[preprint,aps,prl,psfig]{revtex}
\tighten
\draft
\begin{document}

\draft

\preprint{27. October 1999}
\title{High Frequency Dynamics of Amorphous Silica}
\author{J\"urgen Horbach, Walter Kob\footnote{Author to whom
correspondence should be addressed to: e-maiL
Walter.Kob@uni-mainz.de}, and Kurt Binder}
\address{Institut f\"ur Physik, Johannes Gutenberg-Universit\"at,
Staudinger Weg 7, D--55099 Mainz, Germany}
\maketitle

\begin{abstract}
We present the results of extensive molecular dynamics computer
simulations in which the high frequency dynamics of silica,
$\nu>0.5$~THz, is investigated in the viscous liquid state as well as
in the glass state.  We characterize the properties of high frequency
sound modes by analyzing $J_l(q,\nu)$ and $J_t(q,\nu)$, the
longitudinal and transverse current correlation function,
respectively.  For wave--vectors $q>0.4$~\AA$^{-1}$ the spectra are
sitting on top of a flat background which is due to multiphonon
excitations.  In the acoustic frequency band, i.e.~for $\nu<20$~THz,
the intensity of $J_l(q,\nu)$ and $J_t(q,\nu)$ in the liquid and the glass 
approximately proportional to temperature, in agreement
with the harmonic approximation.  In contrast to this, strong
deviations from a linear scaling are found for $\nu>20$~THz.  The
dynamic structure factor $S(q,\nu)$ exhibits for $q>0.23$~\AA$^{-1}$ a
boson peak which is located nearly independent of $q$ around $1.7$~THz.
We show that the low frequency part of the boson peak is
mainly due to the elastic scattering of transverse acoustic modes with
frequencies around 1~THz. The strength of this scattering depends on
$q$ and is largest around $q=1.7$~\AA$^{-1}$, the location of the first
sharp diffraction peak in the static structure factor.  By studying
$S(q,\nu)$ for different system sizes we show that strong finite size
effects are present in the low frequency part of the boson peak in that
for small systems part of its intensity is missing. We discuss the
consequences of these finite size effects for the structural
relaxation.
\end{abstract}


\section{Introduction}
\label{sec1}

The investigation of the vibrational dynamics of supercooled liquids
and glasses is a challenging task since these systems do not have the
property of translational invariance as it is the case for crystals. Of
special interest is the region of intermediate wave--vectors at which
collective excitations, i.e.~longitudinal and transverse sound waves,
begin to experience strongly the structural disorder. Molecular
dynamics (MD) simulations are well suited to study vibrational features
at these intermediate wave--vectors, say with magnitude
$q\geq 0.1$~\AA$^{-1}$, corresponding to frequencies $\nu$ in the THz
band. This paper is concerned with the simulation of the vibrational
dynamics of amorphous silica, which is the prototype of a so--called
strong glassformer \cite{angell85}.  Its structure exhibits a
medium--range order in that it forms a disordered network of
SiO$_4$--tetrahedra leading to a peak in the static structure factor
around $q=1.6$~\AA$^{-1}$ \cite{price87}.  In recent years the high
frequency dynamics of silica has been the subject of an intense debate
because its Raman and neutron scattering spectra
\cite{winterling75,foret96} show a so--called boson peak around
$1$~THz, which is also present in many other glassformers but normally
gives a less intense contribution to the spectra than in silica.
This feature appears also in experiments as an excess over the Debye
density of states or, equivalently, over Debye's $T^3$--law in the
specific heat around the temperature $T=10$~K
\cite{zeller71,buchenau86}.

Recently it has been shown by means of an inelastic X--ray scattering
experiment by Benassi {\it et al.}~\cite{benassi96} that in silica
propagating longitudinal sound modes persist up to $0.35$~\AA$^{-1}$,
which corresponds to frequencies well above the location of the boson
peak.  Therefore Benassi {\it et al.}~argued that the boson peak has
its origin in these propagating sound modes. In contrast to this
suggestion, Vacher {\it et al.}~\cite{vacher97,vacher98,rat99} found
evidence from the available spectroscopic data that the boson peak is
due to the strong scattering of acoustic modes by the disorder thus
regarding the boson peak as a manifestation of these strongly scattered
modes.

Simple models have been suggested to explain the origin of the boson
peak. From the soft potential model \cite{karpov83,buchenau92} the idea
is put forward that anharmonic localized vibrations coexist with
propagating high frequency sound modes in the frequency range around
the location of the boson peak.  In the case of silica these anharmonic
soft modes have been related to coupled SiO$_4$-tetrahedra librations
\cite{buchenau86}.  Wischnewski {\it et al.}~\cite{wischnewski98} have
analyzed their neutron scattering data of silica within the soft
potential model, and have concluded that the sound waves are indeed
scattered from such local vibrational modes below $1$~THz, whereas above
this frequency static Rayleigh scattering from the atomic disorder
takes place.  Schirmacher {\it et al.}~\cite{schirmacher98} have
studied a system of coupled harmonic oscillators with a random
distribution of force constants. In this model they have found an
excess over the Debye behavior in the density of states which they have
interpreted as an analogon to the boson peak feature in real structural
glasses. In agreement with this model, Sokolov \cite{sokolov99} proposed
that the boson peak is related to the strong scattering of acoustic
like vibrations by fluctuations of the elastic constants.

A feature which shares many properties with the boson peak is also
found within the mode--coupling theory (MCT) of the glass transition:
In the ideal glass state where all particles are trapped in the cages
formed by their neighbors, the spectrum of the density--density
correlation function is a superposition of harmonic oscillator spectra
which is due to the variety of cages in which the particles are trapped
\cite{goetze84}.  It is remarkable in this context that the whole light
scattering spectra of glycerol, including the boson peak, have been
successfully described within a schematic MCT model \cite{franosch96},
and very recently G\"otze and Mayr have shown that deep inside the
glass state, i.e. at temperatures well below the MCT temperature $T_c$,
the theory predicts dynamical features which are very reminiscent to
the boson peak~\cite{goetze_draft}.

In the last three years MD simulations tried to give insight into the
vibrational dynamics of silica
\cite{tara97a,tara99a,tara99b,guillot97,horbach98a,dellanna98}, and
other network forming glasses like ZnCl$_2$ \cite{ribeiro98}. Most of
these investigations have analyzed the dynamics within the harmonic
approximation, i.e. by determining the eigenvalues and eigenvectors from the
diagonalized dynamical matrix. Although the full information of the
vibrational part of the dynamics is given by the eigenmodes (of course
only within the harmonic approximation) the origin of the boson peak
remains a puzzle. One reason for this is the smallness of the system
sizes ($20$--$40$~\AA) which have been used in the forementioned
studies, which has the effect that parts of the boson peak are missing
(see below).  A second reason is the difficulty of analyzing the boson
peak feature in terms of eigenmodes since, as we will discuss in detail
below, in the case of silica the coupling of modes with different $q$
is essential for this feature.

In order to avoid these problems we use in the present work a large
system size and calculate the exact current and density correlation
functions in order to investigate their dependence on wave--vector $q$ and
frequency $\nu$.  Therefore, we are able to study the temperature
dependence of the high frequency dynamics of silica in the liquid state
as well as in the glass state. Moreover, we are able to give insight
into the relationship between the vibrational dynamics and structural
relaxation in the silica melt.  The rest of the paper is organized as
follows: In the next section we give an overview of the main
computational details.  In Sec.~\ref{sec3} we discuss the vibrational
dynamics of our silica model by means of the current and density
correlation functions. In Sec.~\ref{sec4} we summarize and discuss the
results.

\section{Model and Details of the Simulations}
\label{sec2}

The model potential we use to describe the interactions between the
ions in silica is the one proposed by van Beest, Kramer, and van Santen
(BKS) \cite{vanbeest91} which has the following functional form:
\begin{equation}
\phi(r)=
\frac{q_{\alpha} q_{\beta} e^2}{r} + 
A_{\alpha \beta} \exp\left(-B_{\alpha \beta}r\right) -
\frac{C_{\alpha \beta}}{r^6}\quad \alpha, \beta \in
[{\rm Si}, {\rm O}].
\label{bkspot}
\end{equation}
Here $r$ is the distance between an ion of type $\alpha$ and an ion of
type $\beta$. The values of the parameters $A_{\alpha \beta}, B_{\alpha
\beta}$ and $C_{\alpha \beta}$ can be found in the original
publication. The Coulombic part of the potential was evaluated by means
of Ewald sums for which further details can be found elsewhere
\cite{horbach99a}. In recent simulations
\cite{horbach99a,horbach99b,vollmayr96,koslowski97,jund99} it has 
been shown that the BKS potential (\ref{bkspot}) reproduces many 
static and dynamic properties of real silica very well and thus it can be
considered as a reliable model for this material.

We have simulated a system with $8016$ ions.  The size of the
simulation box was fixed to $48.365$~\AA~corresponding to a density of
$2.37$~${\rm g}/{\rm cm}^3$. Thus, the smallest wave--vector of our
simulation has magnitude $q=0.13$~\AA$^{-1}$.  In order to study
finite size effects we have done also simulations for smaller systems,
and the details of these simulations are given below.  In the following
we will investigate the fully equilibrated liquid state at $T=6100$~K,
$3760$~K, and $2750$~K and the glass state at $T=1670$~K, $1050$~K, and
$300$~K. In the liquid state we equilibrated the system first at each temperature
in the NVT ensemble at each temperature, and after that we started
microcanonical simulations by means of the velocity form of the Verlet
algorithm.  During the equilibration the temperature was kept constant
by using a stochastic collision algorithm. The time step we used was
1.6~fs, and in order to improve the statistics we simulated at
each temperature two independent runs.  At $T=2750$~K the length of the
equilibration runs was 13 million time steps followed by the
microcanonical production runs over 12 million time steps, which
corresponds to a real time of 20~ns. During the two production runs we
have stored on a linear time scale 30 configurations each which have
subsequently been used as the starting configuration of a new
simulation for investigating the high frequency dynamics.  We mention
that the pressure at $T=2750$~K is around 0.9~GPa.  Further details on
the simulation of the liquid state can be found elsewhere
\cite{horbach99a}.  The starting--point for producing the glass state
were two equilibrated configurations at $T=2900$~K at which the
equilibration time was 4 million time steps (6.5~ns real time). By
coupling the system to an external heat bath the temperature was then
decreased linearly in time within one million time steps to 0~K. This
corresponds to a cooling rate of about $1.8 \cdot 10^{12}$~${\rm
K}/{\rm s}$.  During the cooling procedure we stored configurations at
the temperatures mentioned above which we used as starting
configurations in order to anneal the system for $5\cdot10^5$ time
steps at constant temperature. Afterwards we propagated the system over
$5\cdot10^5$ time steps in the microcanonical ensemble and stored
configurations every $10^5$ time steps. Thus at the end we had at each
of the three temperatures in the glass state 22 starting configurations
for the investigation of the high frequency dynamics.  The pressure for
our glass structures is 0.52~GPa at $T=300$~K, 0.69~GPa at $T=1050$~K,
and 0.8~GPa at $T=1670$~K.

In this paper we are mainly interested in frequency dependent
correlation functions. Therefore time Fourier transformations have to
be calculated which we have done by means of the Wiener--Khinchin
theorem. It says that the Fourier transformation of a correlation
function $C(t)=\left< x(t) x(0) \right>$ ($x(t)$: density, longitudinal
current, transverse current) is given by the power spectrum $Z(\nu)=
\left| a(\nu) \right|^2$ where $a(\nu)$ denotes the Fourier transform
of the time series $x(t)$. The time series were transformed via fast
Fourier transformation whereby we applied a Welch window function
\cite{num_rec}. Usually we have calculated the time series for the
density and the currents over 8192 time steps ($13.4$~ps real time) by
using the aforementioned starting configurations. This results in a
frequency resolution of about $0.1$~THz.  The reliability of the
Fourier transformation was tested by calculating also time series over
16384 time steps and in these test cases we have found indeed identical
spectra, at least for $\nu>0.3$~THz.

\section{Results}
\label{sec3}

\subsection{Current correlations}
\label{sec3.1}

In this section we analyze the vibrational features of our
silica model by means of the longitudinal and transverse current
correlation function $J_l(q,\nu)$ and $J_t(q,\nu)$, respectively, which
depend on the magnitude of the wave--vector ${\bf q}$ and the frequency
$\nu$. These are defined as \cite{boon}
\begin{equation}
 J_{\alpha} (q,\nu) = \frac{1}{N} \int_{-\infty}^{\infty} \; dt \;
	      \exp\left(i 2 \pi \nu t\right) \;
              \left< {\bf j}_{\alpha} ({\bf q},t) \cdot 
		     {\bf j}_{\alpha} (-{\bf q},0)
              \right>
\end{equation}
where the longitudinal part ($\alpha = l$) and the transverse part
($\alpha = t$) of the total current
\begin{equation}
  {\bf j} ({\bf q},t) = \sum_k \dot{{\bf r}}_k (t) 
       \exp \left( i {\bf q} \cdot {\bf r}_k (t) \right)
\end{equation}
are given by
\begin{eqnarray}
  {\bf j}_l ({\bf q},t) & = & \frac{{\bf q} {\bf q}
				    \cdot {\bf j}({\bf q},t)}
                                    {q^2} , \\
  {\bf j}_t ({\bf q},t) & = & {\bf j}({\bf q},t) - 
	  \frac{{\bf q} {\bf q} \cdot {\bf j}({\bf q},t)}{q^2} . 
\end{eqnarray}

\begin{figure}[h]
\psfig{file=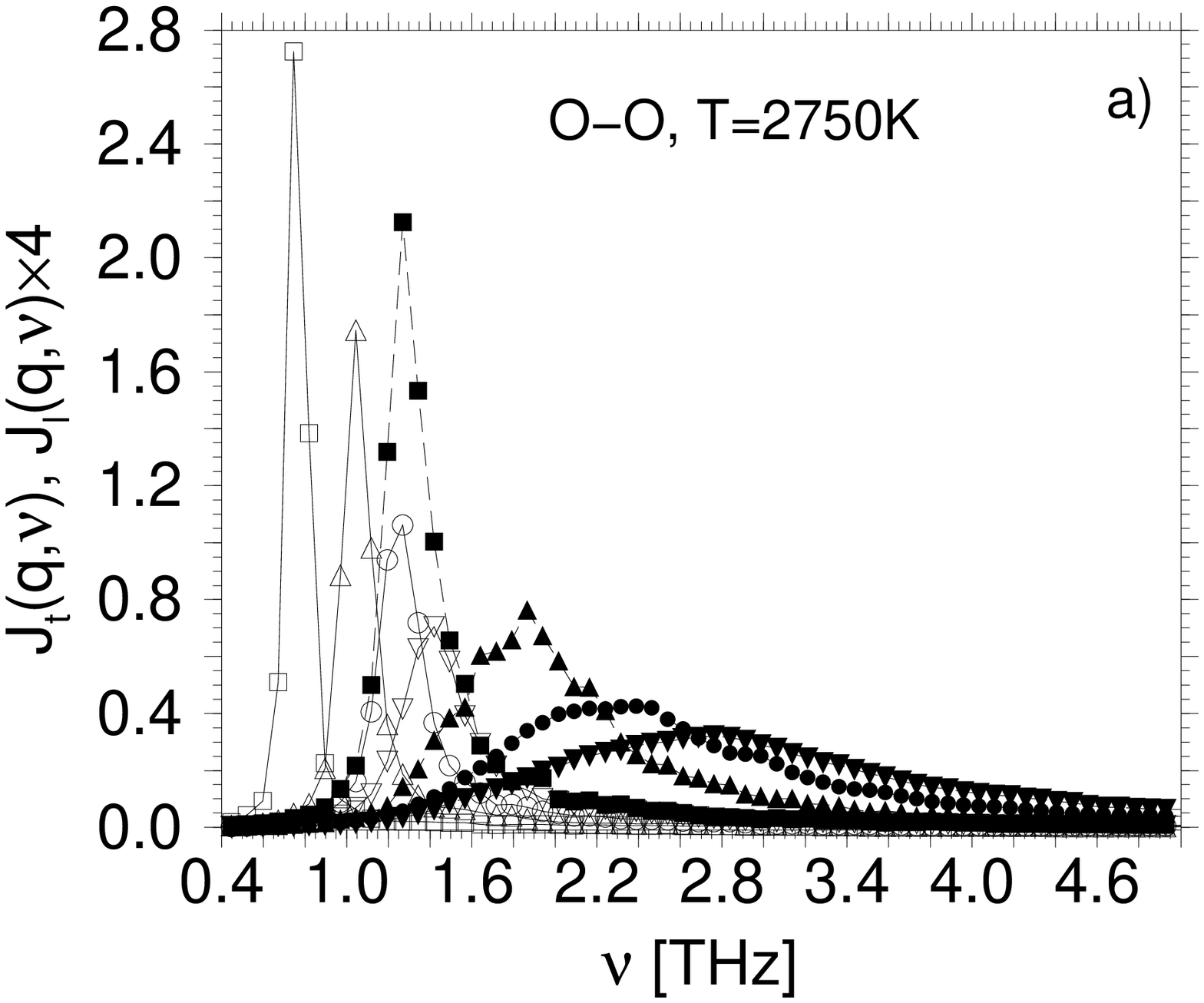,width=13cm,height=9.5cm}
\psfig{file=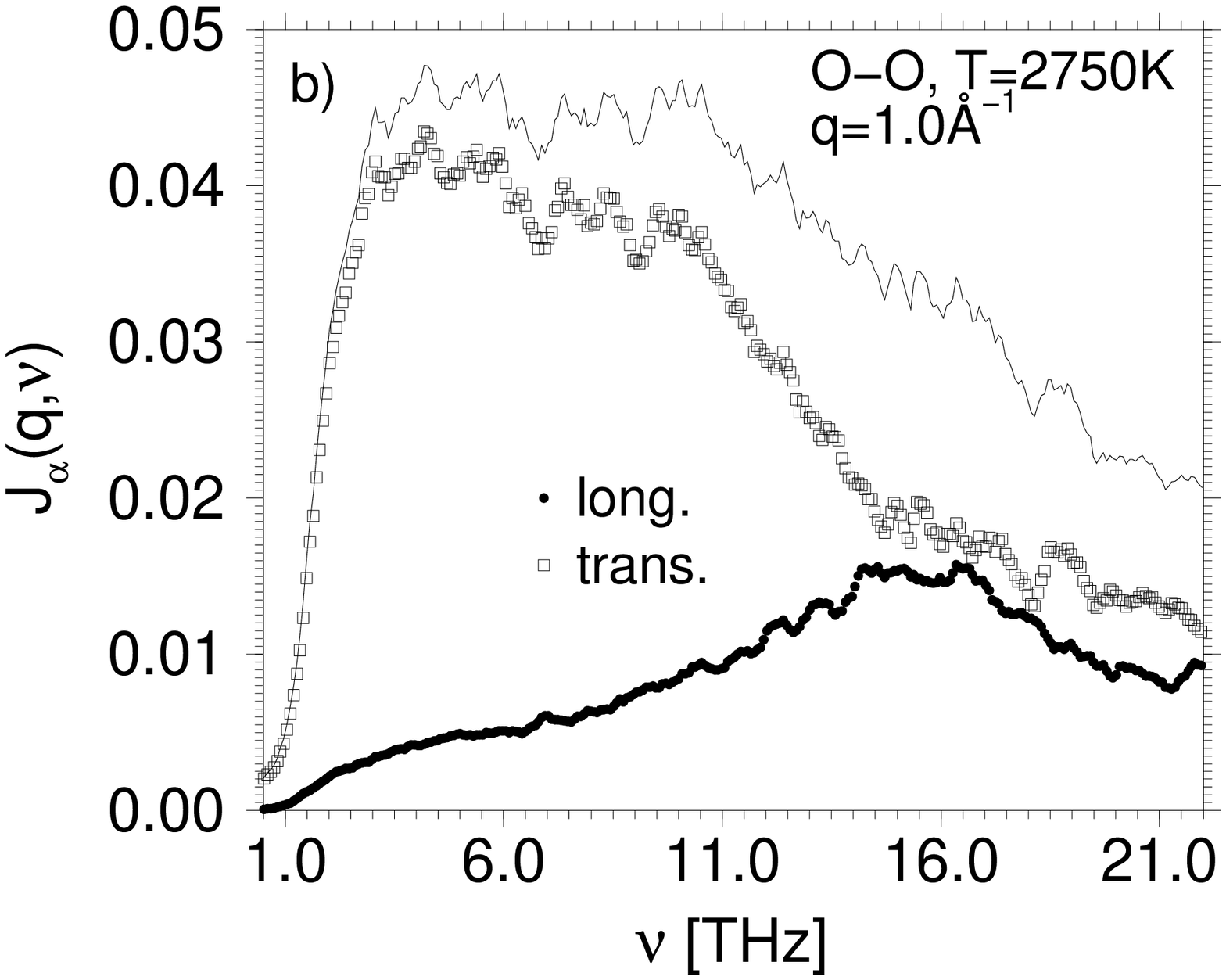,width=13cm,height=9.5cm}
\caption{Longitudinal and transverse current correlation functions
         (filled and open symbols, respectively)
	 for the oxygen--oxygen correlations at the temperature
	 $T=2750$~K. 
         a) The peaks moving to higher frequencies correspond to 
            $q=0.13$~\AA$^{-1}$, $q=0.18$~\AA$^{-1}$, $q=0.23$~\AA$^{-1}$, and 
	    $q=0.26$~\AA$^{-1}$ for the longitudinal and transverse functions,
            respectively. Note that the curves for $J_l(q,\nu)$ are multiplied
            by a factor of $4$.
	 b) $q=1.0$~\AA$^{-1}$. The solid line corresponds
	    to the sum of $J_l$ and $J_t$.}
\label{fig1}
\end{figure}

Fig.~\ref{fig1} shows $J_l(q,\nu)$ and $J_t(q,\nu)$ for different
values of $q$ up to $1.0$~\AA$^{-1}$ at the temperature $T=2750$~K. We
have plotted only the functions for the oxygen--oxygen correlations
because the silicon--silicon and the silicon--oxygen correlations
exhibit qualitatively the same behavior, which is reasonable for such
small wave--vectors. Note that even at the relatively high temperature
$T=2750$~K, our SiO$_2$ model is quite viscous having a viscosity of
about $380$~P, and moreover, that this temperature is well below the
critical temperature of mode--coupling theory, which is at $3330$~K
\cite{horbach99a}. Thus within the framework of MCT we are indeed probing the
system deep in the glass regime.
In Fig.~\ref{fig1}a we show $J_l(q,\nu)$ and
$J_t(q,\nu)$ in the frequency range between 0.4 and 1.6 THz for the
four lowest $q$ values of our simulation, $q=0.13$~\AA$^{-1}$,
$0.18$~\AA$^{-1}$, $0.23$~\AA$^{-1}$, and $0.26$~\AA$^{-1}$. At
$q=0.13$~\AA$^{-1}$ we recognize that there are two peaks,
corresponding to the longitudinal and the transverse part of the
current, which are well separated from each other. For increasing
wave--vectors these peaks move to higher frequencies whereby their
width becomes so large that they overlap more and more with each
other.  In the following we call the excitations corresponding to these
peaks high frequency longitudinal acoustic (LA) modes and high
frequency transverse acoustic (TA) modes, respectively. From the figure
we see that the TA excitations give the most important contribution to
the current spectra in that their amplitude is about a factor $6$--$8$
higher than that of the LA excitations.  In the wave--vector range in
which the LA and TA modes hybridize one would expect that plane waves
are no longer eigenmodes, and in the simulation of Taraskin and Elliott
it has indeed been shown explicitly that a longitudinal or transverse
plane wave with a $q$ value around $0.2$~\AA$^{-1}$ decays into a final
state which can be characterized as a superposition of plane waves with
different wave--vectors and polarizations, but with the same frequency
\cite{tara99a}.

\begin{figure}[h]
\psfig{file=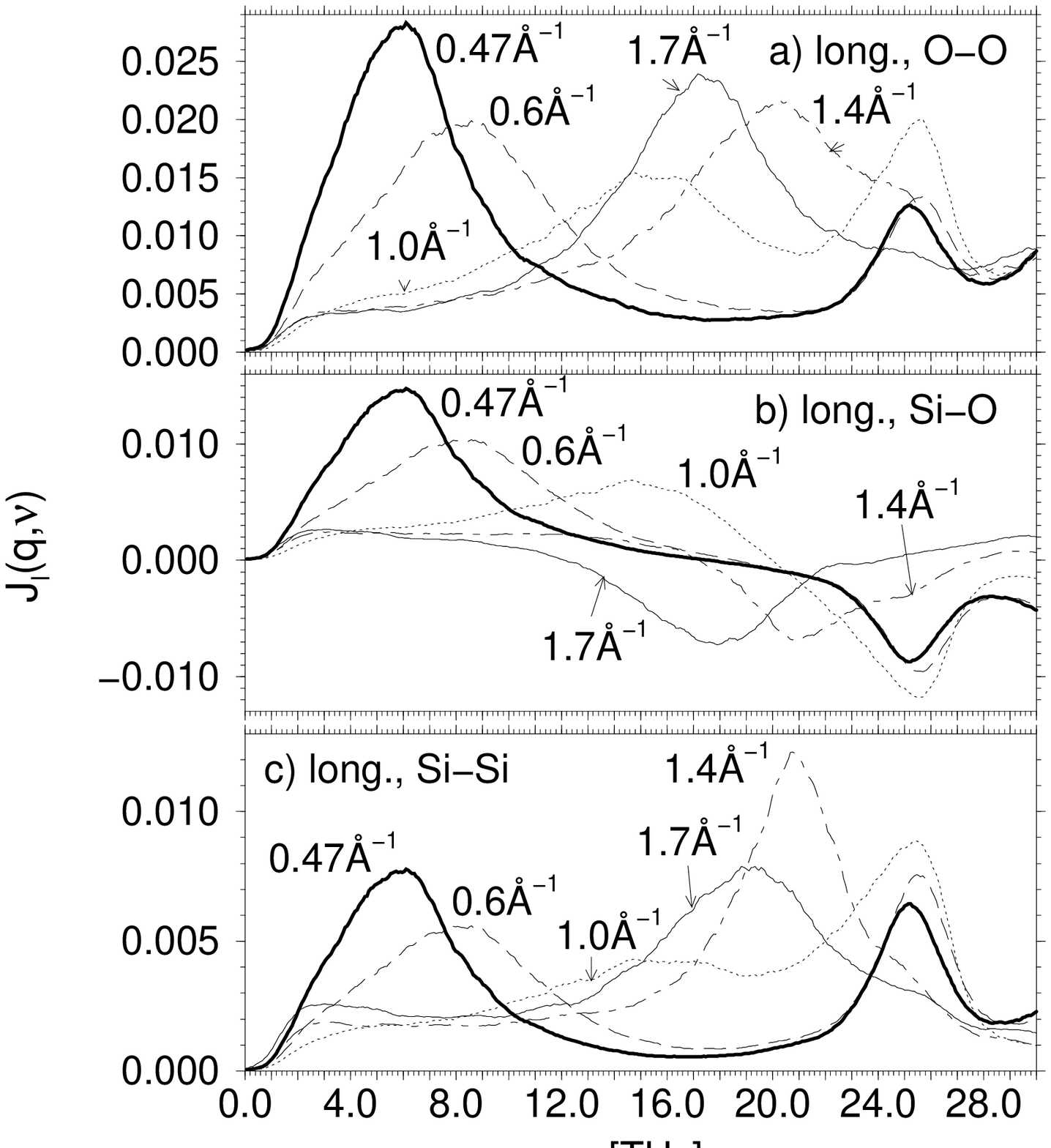,width=11cm,height=9.5cm}
\vspace*{18mm}
\par
\psfig{file=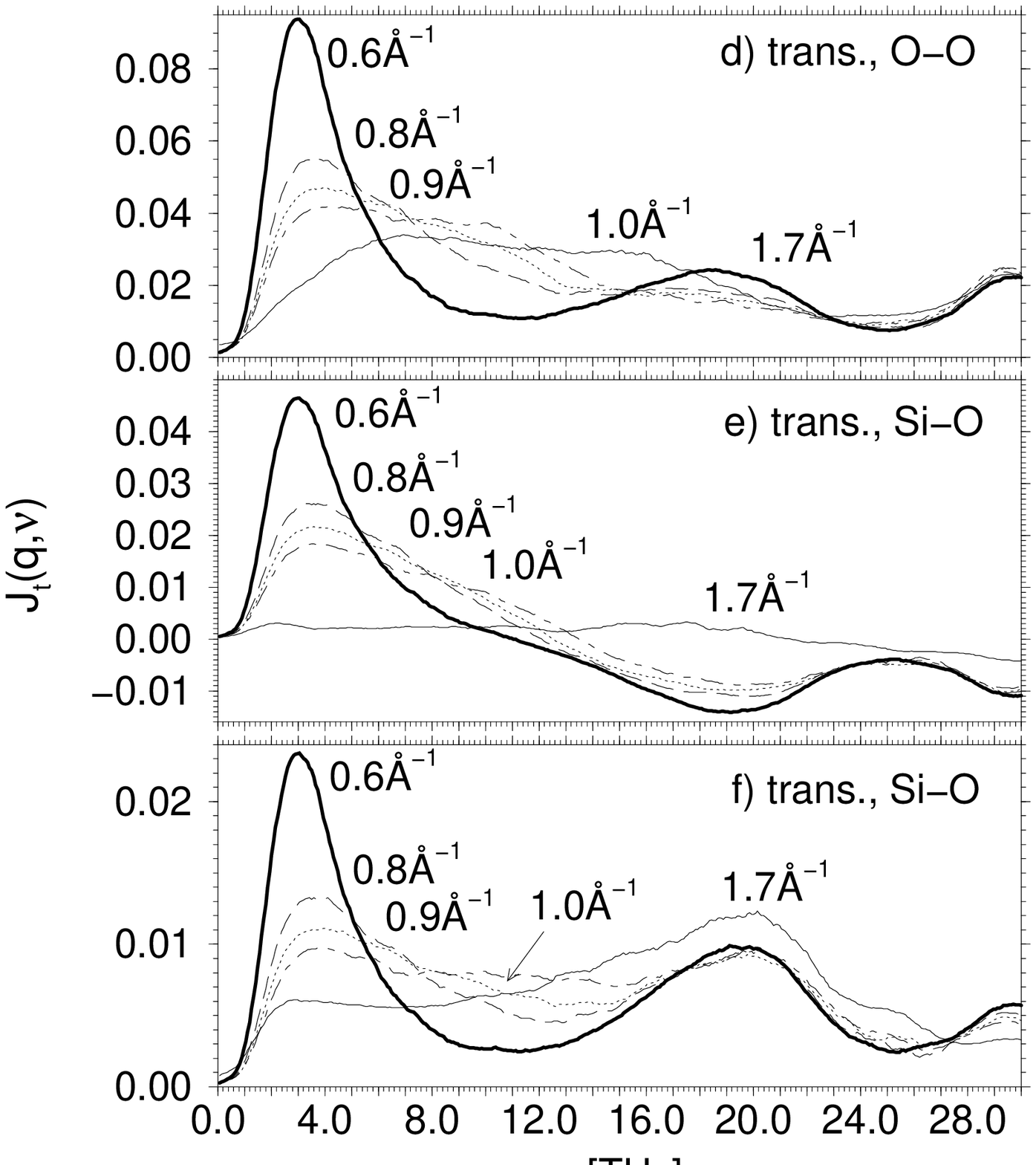,width=11cm,height=9.5cm}
\vspace*{5mm}
\par
\caption{$J_l(q,\nu)$ and $J_t(q,\nu)$ for various wave--vectors at 
         $T=2750$~K, 
         a) $J_l$ for O--O, b) $J_l$ for Si--O,
         c) $J_l$ for Si--Si, d) $J_t$ for O--O,
         e) $J_t$ for Si--O, f) $J_t$ for Si--Si.}
\label{fignew2}
\end{figure}

At $q=1.0$~\AA$^{-1}$ (Fig.~\ref{fig1}b) the current correlation
functions are qualitatively different from those discussed so far at
lower $q$:  In the transverse part one observes a plateau between 3 and
11~THz, and in the longitudinal part the LA peak around 16~THz seems to
be sitting on top of a flat background.  In order to describe in more
detail the change in the shape of the spectra that occurs at
intermediate values of $q$, we show in Fig.~\ref{fignew2} $J_l(q,\nu)$
and $J_t(q,\nu)$ for the O--O, Si--O, and Si--Si correlations for $q$
up to $1.7$~\AA$^{-1}$. At $q=0.47$~\AA$^{-1}$ we observe in
$J_l(q,\nu)$ for the O--O correlations, apart from the LA peak around
$6$~THz, a peak around $\nu=26$~THz corresponding to an optical
excitation.  Moreover, the intensity of the whole spectrum seems to be
enhanced in that the LA and the optical peak sit on top of a flat
background.  If $q$ is increased to $1.4$~\AA$^{-1}$ the LA peak moves
to larger frequencies whereby a shoulder around $\nu=2$~THz gets more
and more pronounced.  Also at $q=1.7$~\AA$^{-1}$ there is a LA peak but
now its position has moved back to $\nu=17$~THz, i.e.~a smaller frequency
than at $q=1.4$~\AA$^{-1}$.  In the case of the Si--O correlations
(Fig.~\ref{fignew2}b) the essential difference to the O--O correlations
is the negative amplitude of the LA peak for $q\ge1.4$~\AA$^{-1}$ which
indicates an antiphase motion of the silicon and oxygen atoms. The
curves for the Si--Si correlations (Fig.~\ref{fignew2}c) show
essentially only one difference compared to those for the O--O
correlations in that the optical band has a higher weight in the
spectrum than the LA excitations. This is due to the fact that the
silicon atoms are bonded stronger in the tetrahedral network than the
oxygen atoms, and thus on small length scales more localized motions
have a higher weight in the case of the silicon atoms which corresponds
to frequencies in the optical band.

Also in the transverse case for the O--O correlations
(Fig.~\ref{fignew2}d) the whole spectrum sits on top of a flat
background. The intensity of the TA peak around $3$~THz decreases with
increasing $q$ whereas there is an increase in the intensity around
9~THz. As a result a broad flat band is obtained for $\nu< 17$~THz.  In
contrast to the O--O correlations, $J_t(q,\nu)$ for the Si--O
correlations (Fig.~\ref{fignew2}e) shows a strong overall decrease of the
intensity if $q$ is increased from $1.0$~\AA$^{-1}$ to
$1.7$~\AA$^{-1}$. This can be easily understood because at
$q=1.7$~\AA$^{-1}$ the current correlation functions measure to a great
extent the single particle motion, and therefore the relative motion of
the silicon and oxygen atoms gives only a small contribution to the
spectra.  The most remarkable feature in $J_t(q,\nu)$ for the Si--Si
correlations is again that, compared to the O--O correlations, the
optical excitation around 20~THz has a larger amplitude than those of
the acoustic band for $q\ge1.0$~\AA$^{-1}$.  The essential result which
is shown in the Fig.~\ref{fignew2} is that for intermediate values of
$q$ the whole spectrum is placed on top of a flat background. A similar
feature has also been found by Mazzacurati {\it et
al.}~\cite{mazzacurati96} in a Lennard--Jones system. These authors
have identified the flat background directly in the spectra and in the
participation ratio which measures the number of particles that
contribute to the eigenmodes at a certain frequency. At the low
frequency edge of the density of states the participation ratio has
values expected for localized modes. Such a behavior of the
participation ratio has also been found in the case of silica
\cite{tara97b}. Mazzacurati {\it et al.}~have explained this behavior
by showing that the eigenvectors for low frequencies can be represented
by a few long--wavelength standing waves plus a random contribution
where the random contribution is seen in the spectrum as the flat
background.  In a phonon picture one can interpret the flat background
as the contribution of multiphonon excitations.

\begin{figure}[h]
\psfig{file=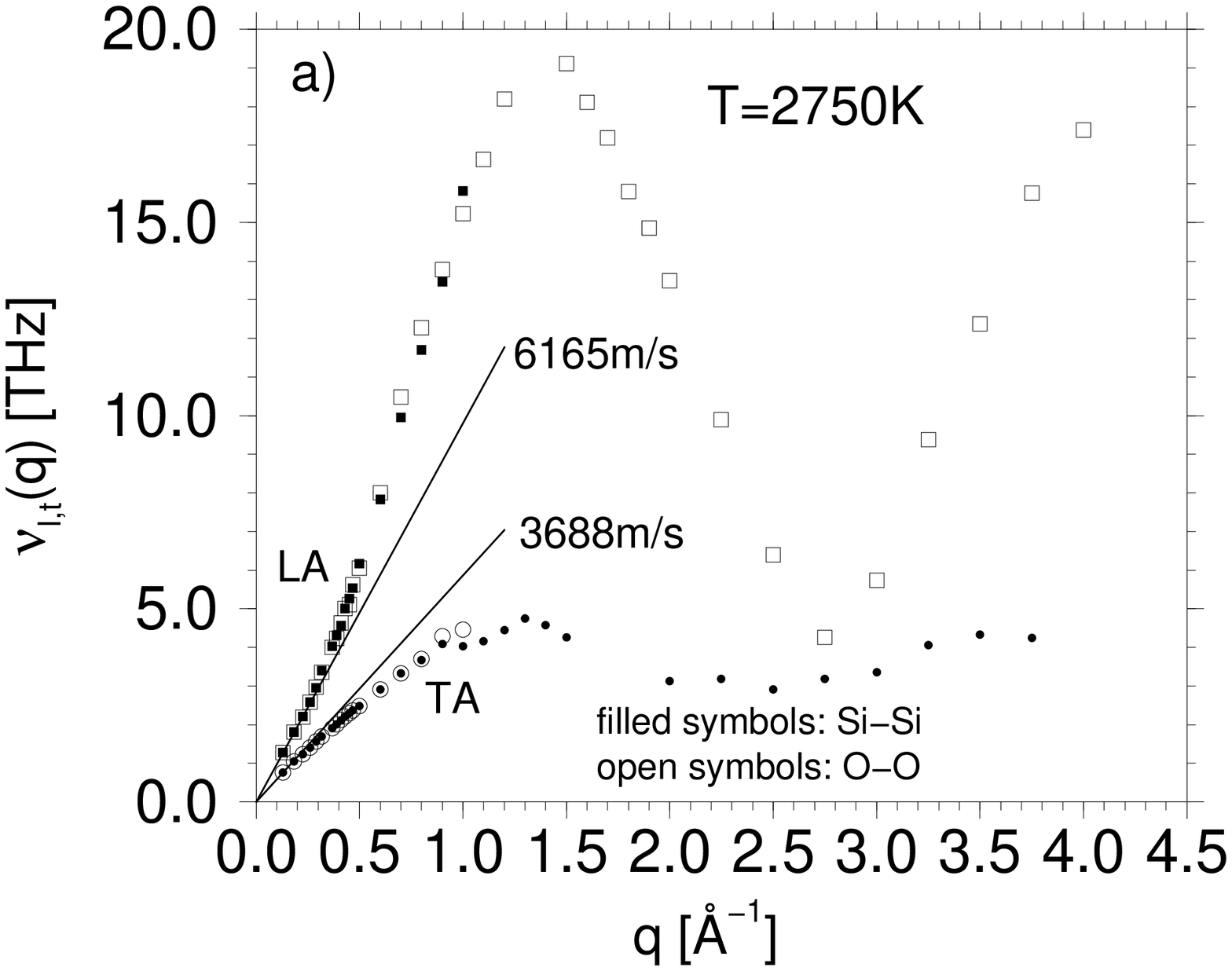,width=13cm,height=9.5cm}
\psfig{file=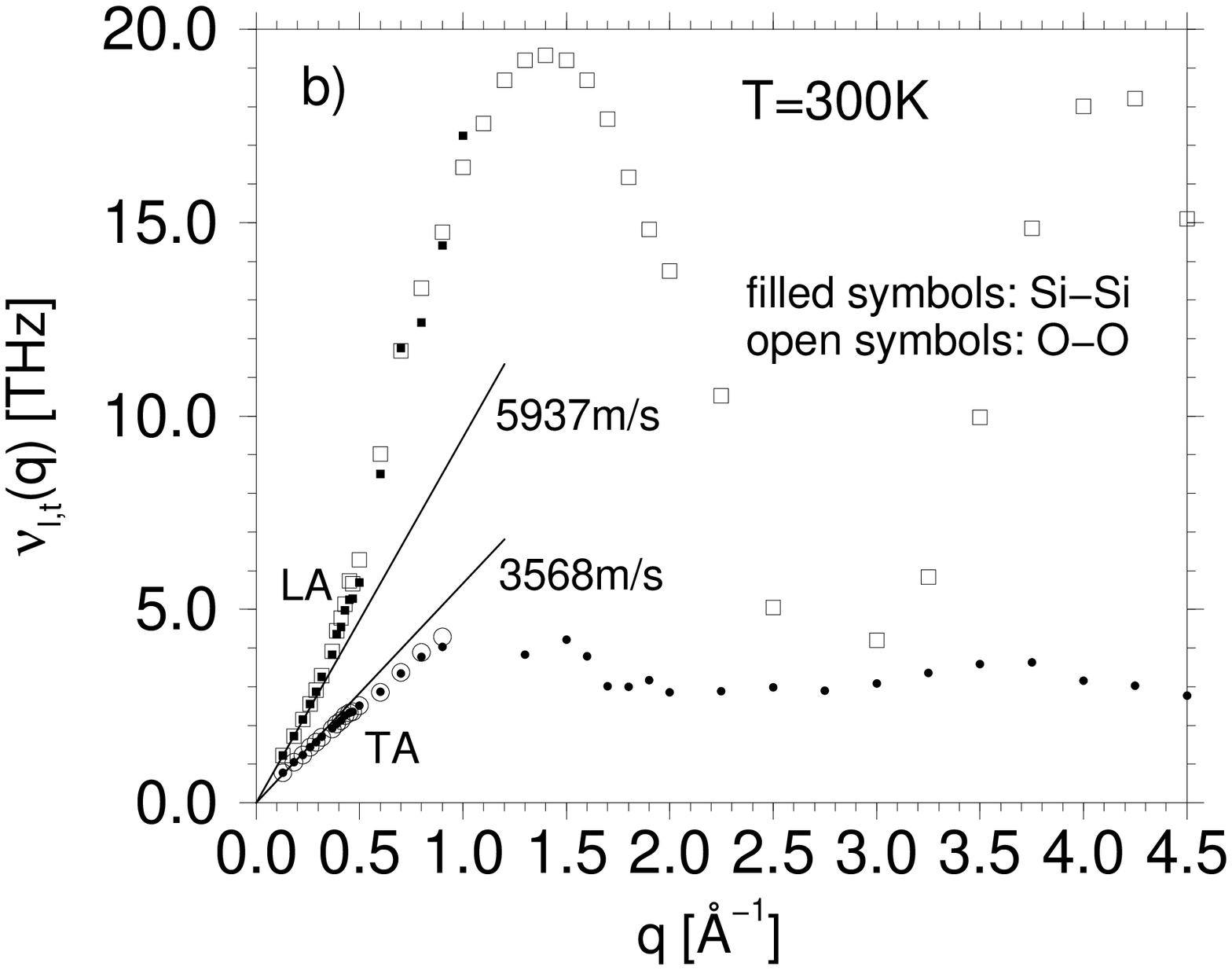,width=13cm,height=9.5cm}
\caption{Peak maximum position $\nu_{l,t}(q)$ for the LA and TA modes
         for the Si--Si correlations (filled symbols) and the 
         O-O correlations (open symbols) at a) $T=2750$~K and 
         b) $T=300$~K. The bold lines are fits with linear laws
         $\nu_{\alpha}=c_{\alpha} q / (2 \pi)$ for which the corresponding
         values for the sound velocities $c_{\alpha}$ are given in the figure.}
\label{fig2}
\end{figure}

By reading off the peak maxima \cite{peak_max} in $J_l(q,\nu)$ and
$J_t(q,\nu)$ corresponding to the longitudinal and transverse acoustic
modes one gets dispersion like branches $\nu_l(q)$ and $\nu_t(q)$ which
are shown in Fig.~\ref{fig2}a for $T=2750$~K and in Fig.~\ref{fig2}b
for $T=300$~K. It is remarkable that $\nu_l(q)$ and $\nu_t(q)$ exhibit
essentially the same behavior in the viscous liquid state ($T=2750$~K)
and the glass state ($T=300$~K). This shows that in this (high)
frequency window there is no relevant difference between a viscous
liquid and a glass which gives support to the idea of Ref.~\cite{goetze_draft}
that in this frequency range the viscous liquid can be treated like a glass.
Furthermore, we note that both functions look very
similar as in simple liquids \cite{boon}:  The longitudinal branch
$\nu_l(q)$ has a periodic structure with a minimum located around
$q_{{\rm m}}=2.8$~\AA$^{-1}$, which is the location of the second sharp
diffraction peak in the static structure factor and which corresponds
to length scales of intratetrahedral distances \cite{horbach99a}.
Thus, $q_{{\rm m}}/2$ can be interpreted, in analogy to crystals, as a
quasi Brillouin zone. The minimum in $\nu_l(q)$ at $q_{{\rm m}}$ can be
easily understood since the particles tend to favor relative
separations of $2 \pi / q_{{\rm m}}$, and therefore, at these
wavelengths it costs a relatively small amount of energy to excite a
collective mode corresponding to a relatively small frequency.  That
the minimum in $\nu_l(q)$ is not observed at $q=1.7$~\AA$^{-1}$, the
location of the {\it first} sharp diffraction peak in the static
structure factor \cite{horbach99a}, is due to the fact that this $q$
value corresponds to length scales of connected SiO$_4$--tetrahedra, a
structural unit which is less stiff than one tetrahedron itself.  The
behavior of $\nu_l(q)$ is in agreement with the findings in a neutron
scattering experiment by Arai {\it et al.}~\cite{arai98}, and was also
found in the computer simulations of Taraskin and Elliott
\cite{tara97b}.  The transverse branch $\nu_t(q)$ becomes rather flat
for $q>0.9$~\AA$^{-1}$ which is an indication of the overdamped
character of the TA excitations at these wave--vectors.

\begin{figure}[h]
\psfig{file=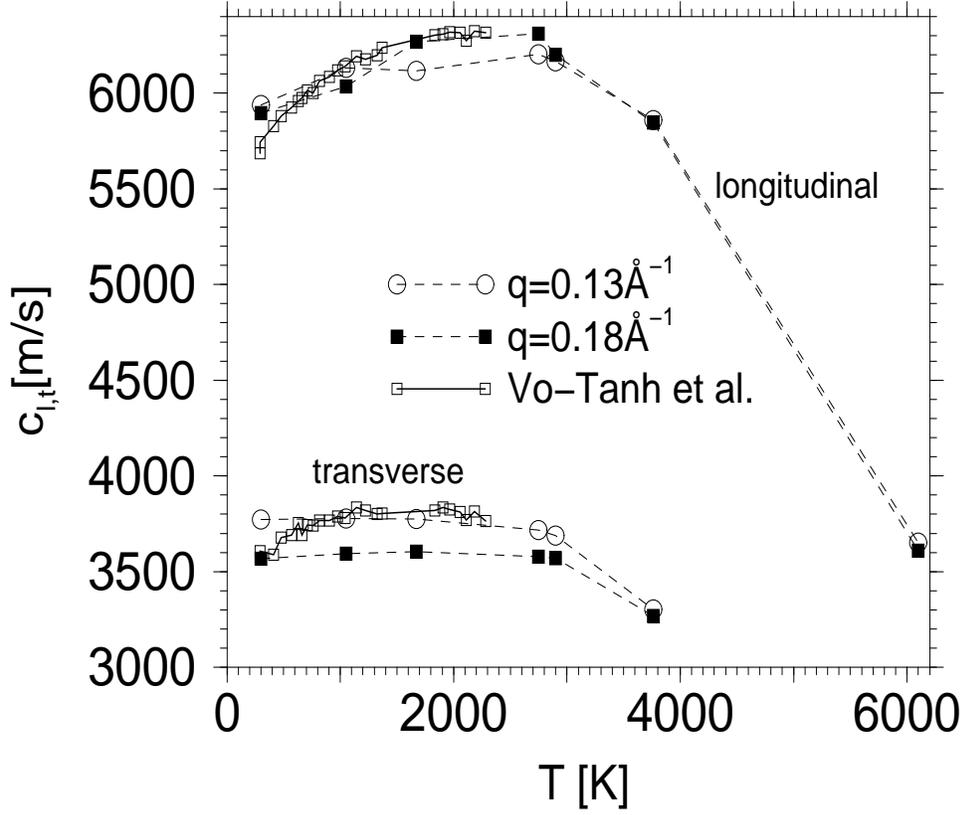,width=13cm,height=10.5cm}
\vspace*{5mm}
\par
\caption{The temperature dependence of the sound velocities $c_{l,t}$
         which are determined from $\nu_{l,t}(q)$ at 
         $q=0.13$~\AA$^{-1}$ (open circles) and 
         $q=0.18$~\AA$^{-1}$ (filled squares). Also included are the 
         experimental data of 
         Vo--Tanh {\it et al.}~\protect\cite{votanh} which we have
         multiplied with the factor $(2.2 / 2.37)^{0.5}$ in order to
         to take into account the different density of our simulation 
         from that of the experiment.}
\label{fig3}
\end{figure}

Also included in Figs.~\ref{fig2}a and \ref{fig2}b are fits of the form
$\nu_{\alpha}(q)=c_{\alpha} q / (2 \pi)$, where $c_l$ and $c_t$ denote
the longitudinal and the transverse high frequency sound velocity,
respectively. The values for $c_{\alpha}$ obtained from these fits are
given in the figures.  We recognize that for $q$ up to around
$0.4$~\AA$^{-1}$ this linear dispersion law holds, which is expected
for propagating sound waves at sufficiently small $q$.  We have
determined the longitudinal and transverse sound velocity for all
temperatures considered by calculating $c_{\alpha}= 2 \pi \nu_{\alpha}
/ q$ for the two lowest $q$ values of our simulation
$q=0.13$~\AA$^{-1}$ and $q=0.18$~\AA$^{-1}$. The sound velocities
obtained in this way are shown in Fig.~\ref{fig3} as a function of
temperature.  Note that $c_{\alpha}$ as determined from
$q=0.13$~\AA$^{-1}$ and from $q=0.18$~\AA$^{-1}$ differ by less than
$7$~\% from each other, which shows that these wave--vectors are small
enough to determine $c_{{\alpha}}$ reliably. From $3760$~K to $6100$~K
the longitudinal sound velocity decreases by about $50$~\%.  No data is
shown for $c_t$ at $6100$~K in the figure because at this temperature
only a peak at $\nu=0$ is observed. This behavior is in agreement with
hydrodynamics which predicts that transverse fluctuations are
transported diffusively and therefore contribute to the spectrum only
with a peak at $\nu=0$.  We have found, however, that even at $T=6100$K
the restoring forces between the particles are large enough to allow
the propagation of TA modes for $q\geq 0.35$\AA$^{-1}$, which can be
inferred by the observation of a crossover from a peak around $\nu=0$
to a peak at finite frequencies in this region of $q$. 
Also included in
the figure are the experimental sound velocities measured by Vo--Tanh
{\it et al.}~\cite{votanh} which are multiplied with the factor
$\sqrt{2.2/2.37}$. This factor takes into account that the density of
our simulation, $2.37 \; {\rm g}/{\rm cm}^3$, is slightly different
from the experimental one, which is $2.2 \; {\rm g}/{\rm cm}^3$. With
this ``correction'' the simulation reproduces the experimental data
very well, both for the longitudinal and the transverse sound
velocities. Note that the experimental data have been obtained by
Vo--Tanh {\it et al.}~by means of light scattering experiments for
values of $q$ of the order $10^{-3}$~\AA$^{-1}$, i.e.~about two orders
of magnitude below the $q$ values of our study.  Since, however, it has
been shown by Benassi {\it et al.}~\cite{benassi96} that at least the
longitudinal sound velocities do not change in this $q$ range,
i.e.~essentially the same value for $c_l$ is measured in Brillouin
scattering experiments and in X--ray scattering up to
$q\approx0.35$~\AA$^{-1}$, it is reasonable to compare the values of
$c_{\alpha}$ from Ref.~\cite{votanh} with our data.

\begin{figure}[h]
\psfig{file=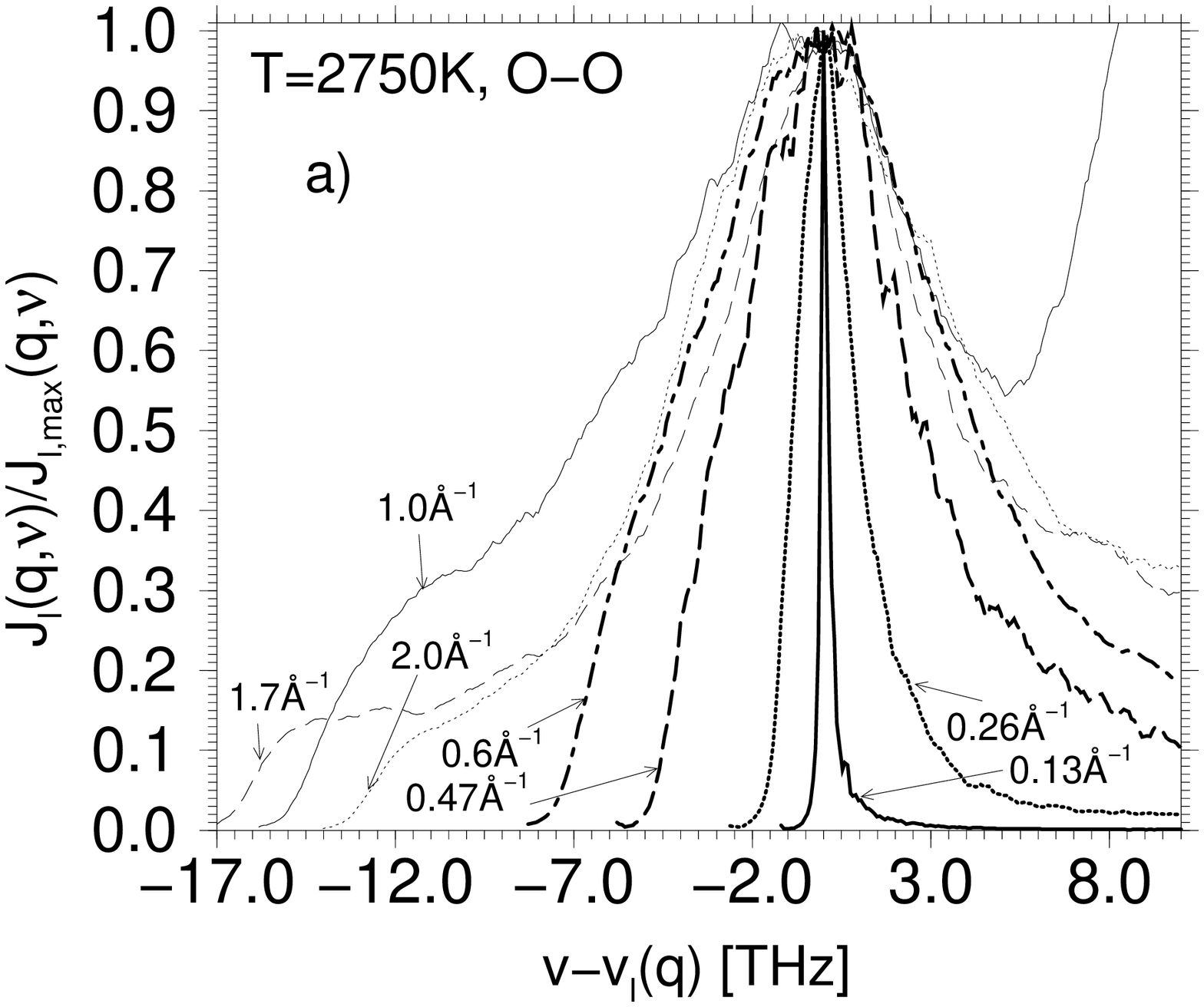,width=13cm,height=9.5cm}
\psfig{file=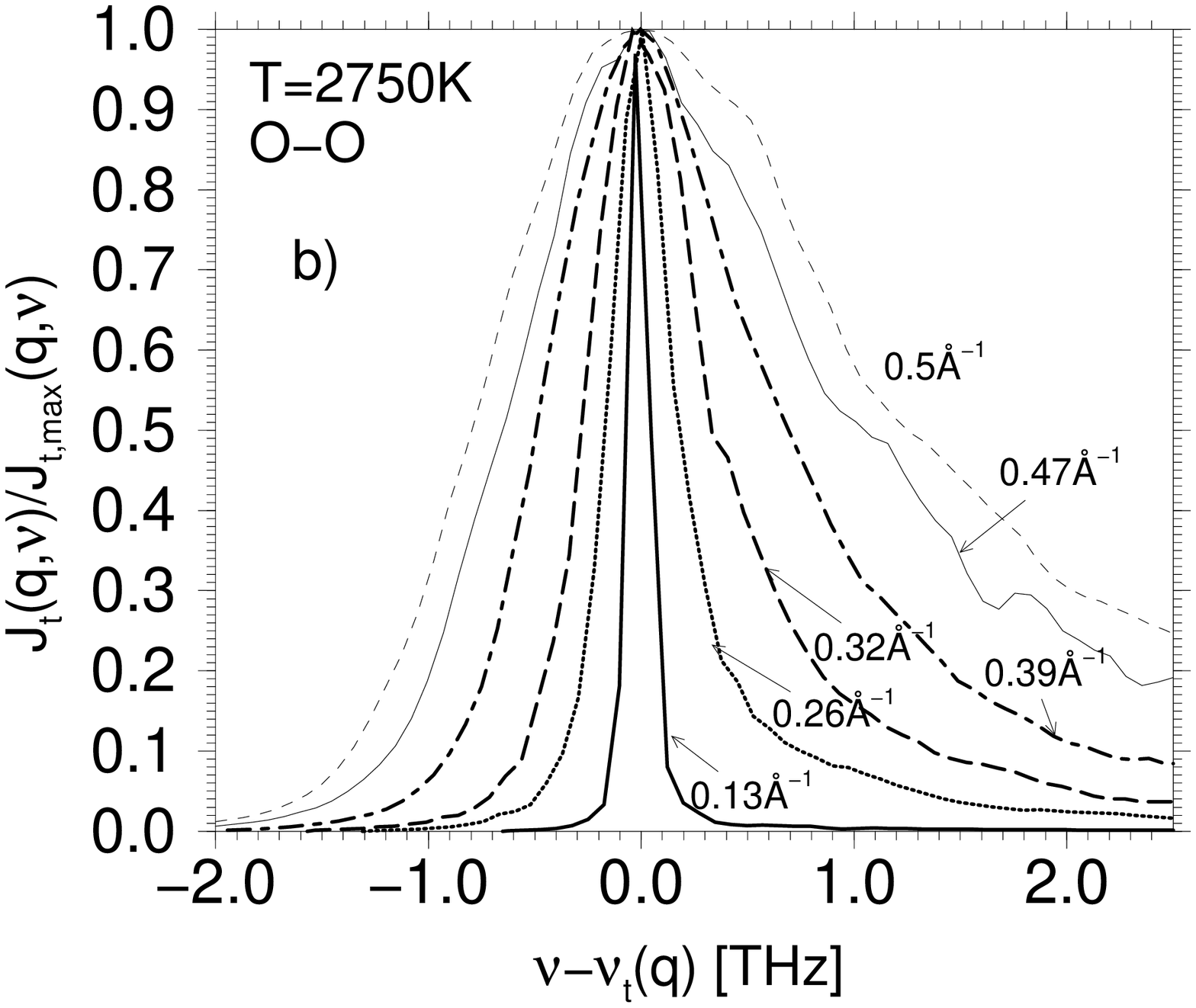,width=13cm,height=9.5cm}
\caption{a) $J_l(q,\nu)/J_{l,{\rm max}}$ versus $\nu-\nu_l(q)$
            at $T=2750$~K for the O--O correlations in the $q$ range
            $0.13$~\AA$^{-1} \; \le q \le \; 2.0$~\AA$^{-1}$.
         b) $J_t(q,\nu)/J_{t,{\rm max}}$ versus $\nu-\nu_t(q)$
            at $T=2750$~K for the O--O correlations in the $q$ range
            $0.13$~\AA$^{-1} \; \le q \le \; 0.5$~\AA$^{-1}$.} 
\label{fig4}
\end{figure}

In order to determine, independent from a model, the width of the peaks
corresponding to the LA and TA excitations we have plotted the ratio
$J_{\alpha}(q,\nu)/J_{\alpha,{\rm max}}(q,\nu)$ versus
$\nu-\nu_{\alpha}(q)$ where $J_{\alpha,{\rm max}}$ denotes the
amplitude of $J_{\alpha}(q,\nu)$ at the maximum frequency
$\nu_{\alpha}(q)$.  The resulting curves are shown in
Fig.~\ref{fig4}a in the longitudinal case for $q$ values up to
$2.0$~\AA$^{-1}$ and in Fig.~\ref{fig4}b in the transverse case for $q$
values up to $0.5$~\AA$^{-1}$ from which we have read off the full
width at half maximum $\Gamma_{\alpha}(q)$.  We recognize from
Fig.~\ref{fig4}a that the broadening in the curves for the longitudinal
part is non--monotonous for $q>1.0$~\AA$^{-1}$ whereas the curves in
the transverse case broaden monotonously, as can be seen in
Fig.~\ref{fig4}b.

\begin{figure}[h]
\psfig{file=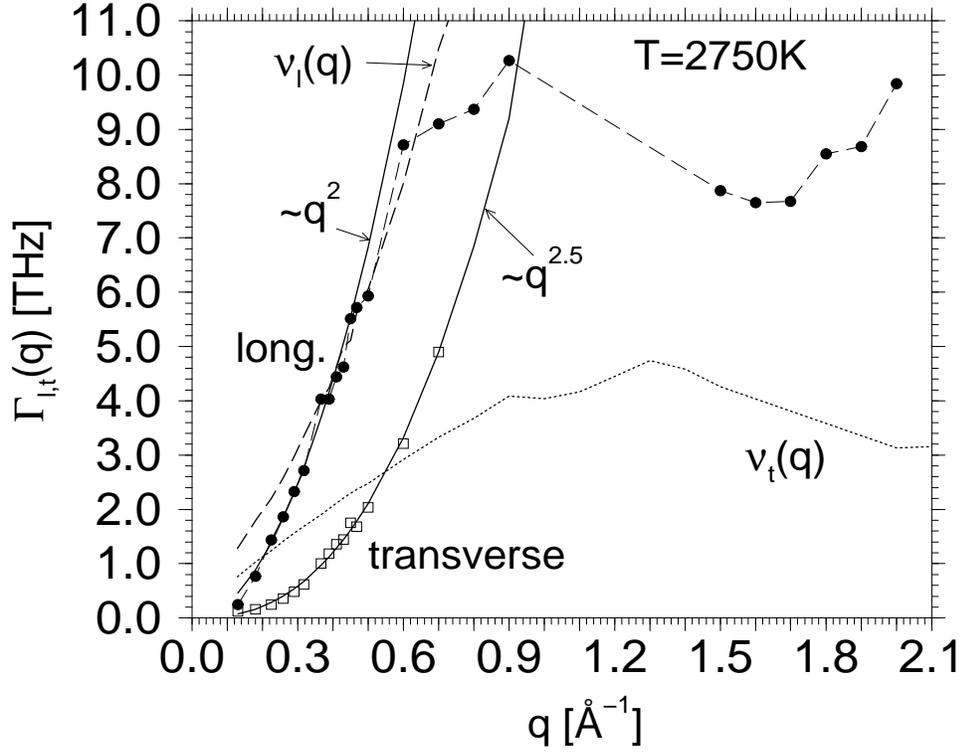,width=13cm,height=9.8cm}
\vspace*{5mm}
\par
\caption{Linewidth $\Gamma_{l,t}(q)$ obtained at $T=2750$~K. The bold 
	 lines are power law fits with exponents $2$ and $2.5$ for
	 $\Gamma_l(q)$ and $\Gamma_t(q)$, respectively. Also included
         are $\nu_l(q)$ (bold dashed line) and $\nu_t(q)$ (bold dotted line).}
\label{fig5}
\end{figure}

The linewidths $\Gamma_{\alpha}(q)$ obtained in this way 
are shown in Fig.~\ref{fig5} for
the LA and TA modes. We recognize from this figure that a quadratic fit
describes the longitudinal half width well in the $q$ interval
$0.18$~\AA$^{-1} \le q \le 0.5$~\AA$^{-1}$.  Such a behavior has also
been found in the experiment by Benassi {\it et al.}~\cite{benassi96}
which was done at $T=1050$~K.  The linear dispersion for $\nu_l(q)$ and
the quadratic law $\Gamma_l(q) \propto q^2$ support the picture that
for $q<0.4$~\AA$^{-1}$ the system behaves like an isotropic elastic
medium with respect to the propagation of the bare LA phonons.  For
$q\ge 0.6$~\AA$^{-1}$ $\Gamma_l(q)$ becomes rather flat up to
$q=1.1$~\AA$^{-1}$. Then this function decreases significantly and
reaches a minimum in the vicinity of the first sharp diffraction peak
at $q=1.6$~\AA$^{-1}$.  This is probably due to the fact that the
strong spatial correlations on the length scale of two connected
SiO$_4$ tetrahedra decreases the damping of the LA excitation.  It is
interesting that in the $q$ range $0.6$~\AA$^{-1} < q < 2.0$~\AA$^{-1}$
the width $\Gamma_l(q)$ is significantly smaller than $\nu_l(q)$ (see
Fig.~\ref{fig5}) which means that the LA excitations cannot be
characterized by a Ioffe--Regel crossover.  
Note that no data points
are available between $1.1$ and $1.4$~\AA$^{-1}$ because the LA peak
overlaps with the optical band in this $q$ range and thus they cannot
be identified uniquely.  

From Fig.~\ref{fig5} we also see that the
transverse peak width $\Gamma_t(q)$ can be described by an
effective power law with exponent $2.5$, $\Gamma_t(q) \propto
q^{2.5}$.  Thus, the TA phonons seems to be damped stronger than
expected from an isotropic elastic medium which would give an exponent
$2$.  In the $q$ region above $0.5$~\AA$^{-1}$ the width $\Gamma_t(q)$
becomes larger than the corresponding location of the maximum of the
peak $\nu_t(q)$.  Therefore, we observe a Ioffe--Regel crossover in the
transverse case where the TA excitations lose their propagative
character and become strongly overdamped.  Note that this is the $q$
region for which $\nu_t(q)$ becomes more or less flat, in contrast to $\nu_l(q)$.

\begin{figure}[h]
\psfig{file=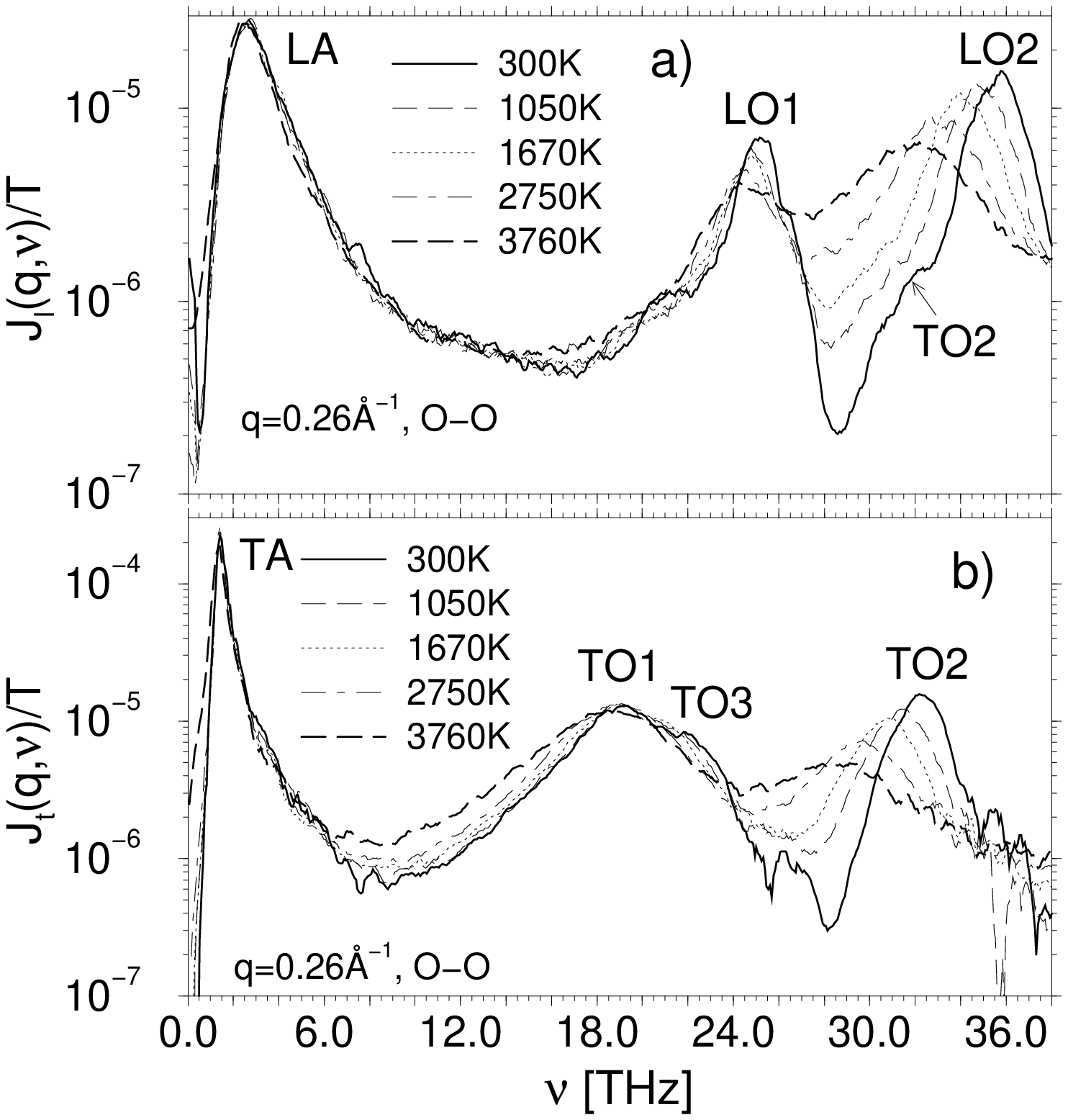,width=11cm,height=9.5cm}
\psfig{file=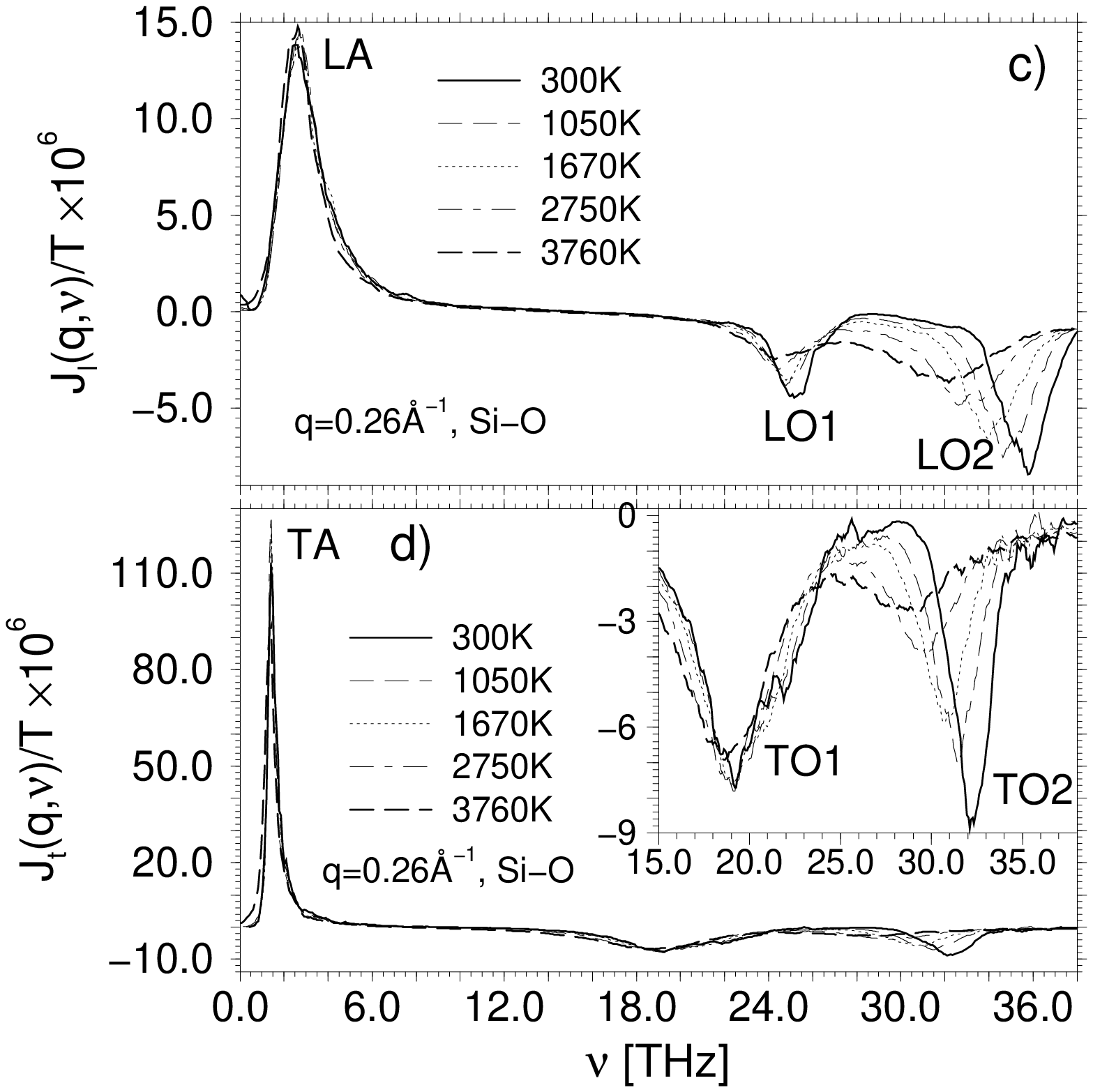,width=11cm,height=9.5cm}
\psfig{file=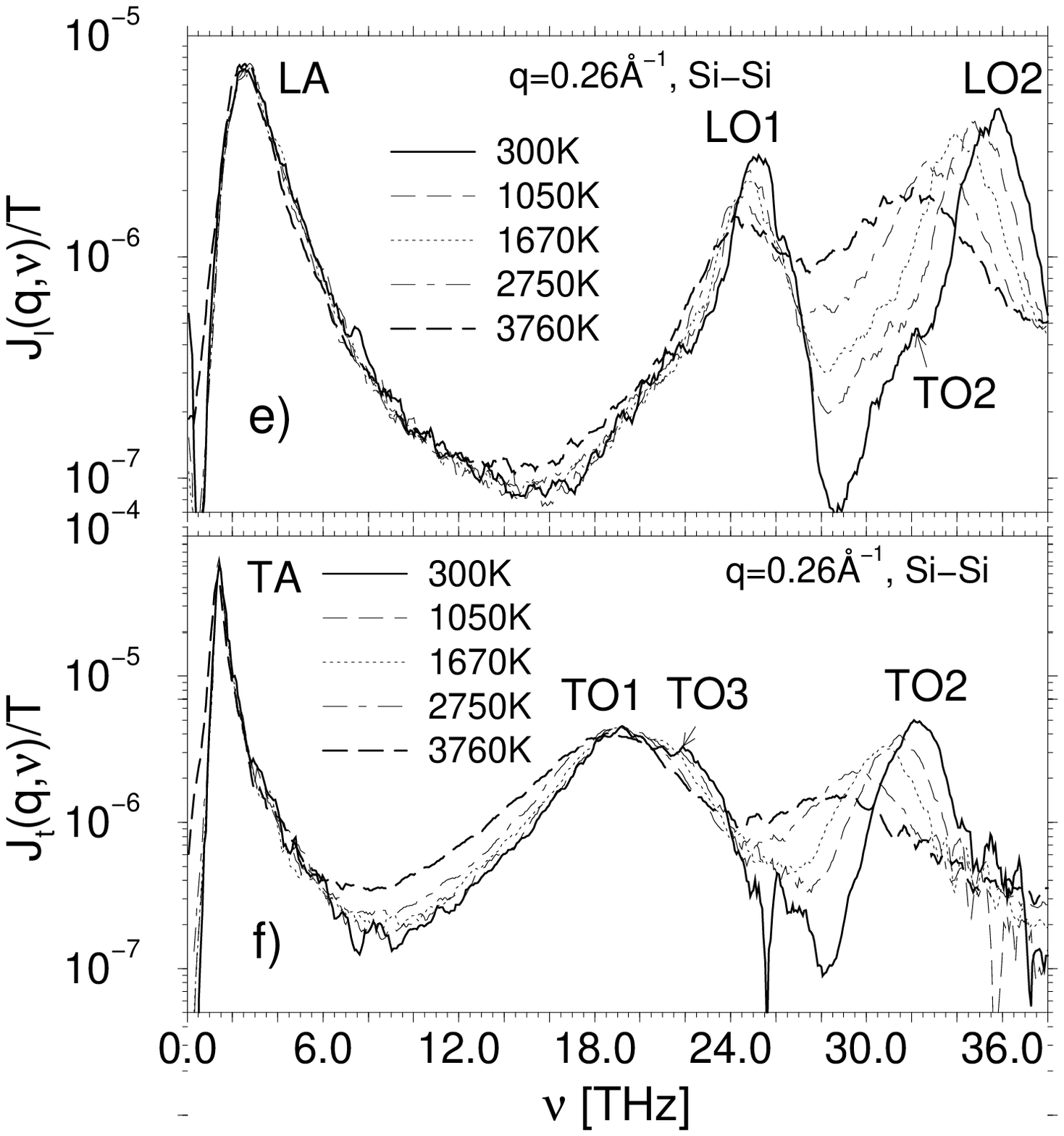,width=11cm,height=9.5cm}
\caption{The current correlation functions scaled with temperature
	 for $q=0.26$~\AA$^{-1}$,  
         a) $J_l(q,\nu)/T$ for the O--O correlations, 
         b) $J_t(q,\nu)/T$ for the O--O correlations,
         c) $J_l(q,\nu)/T$ for the Si--O correlations,
         d) $J_t(q,\nu)/T$ for the Si--O correlations,
         e) $J_l(q,\nu)/T$ for the Si--Si correlations,
         f) $J_t(q,\nu)/T$ for the Si--Si correlations.}
\label{fig6}
\end{figure}

\begin{figure}[h]
\psfig{file=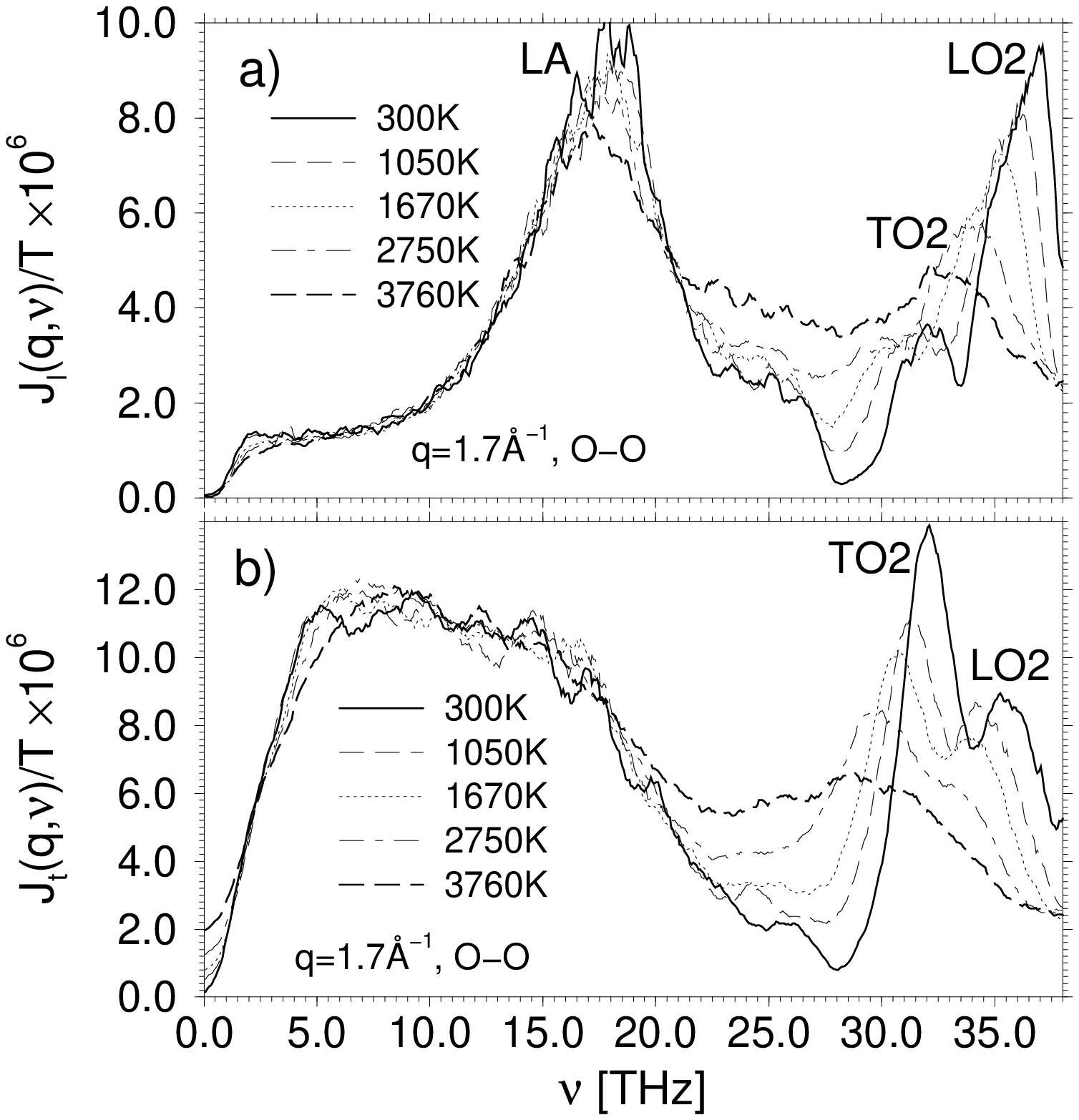,width=9cm,height=7.5cm}
\psfig{file=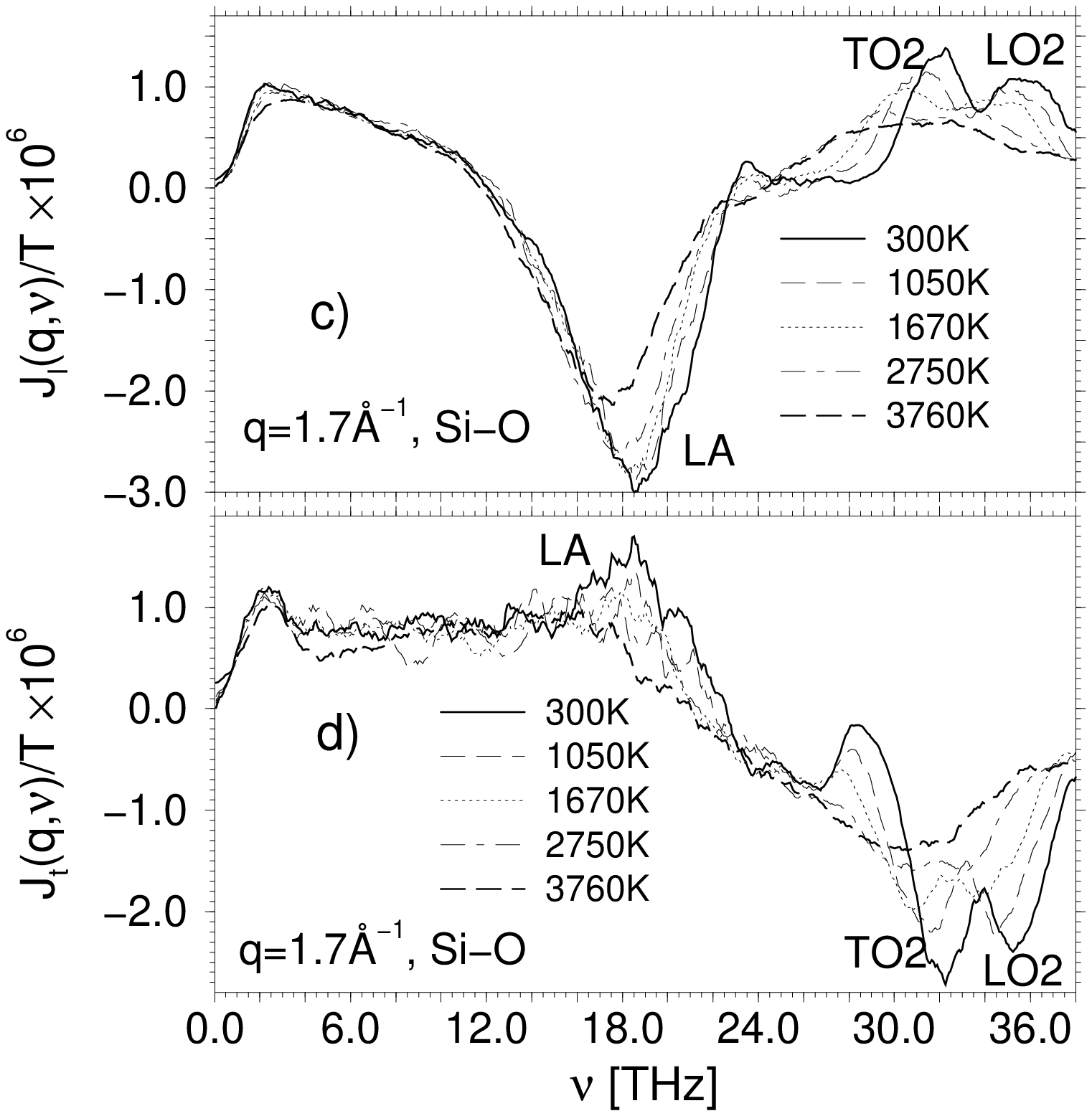,width=9cm,height=7.5cm}
\psfig{file=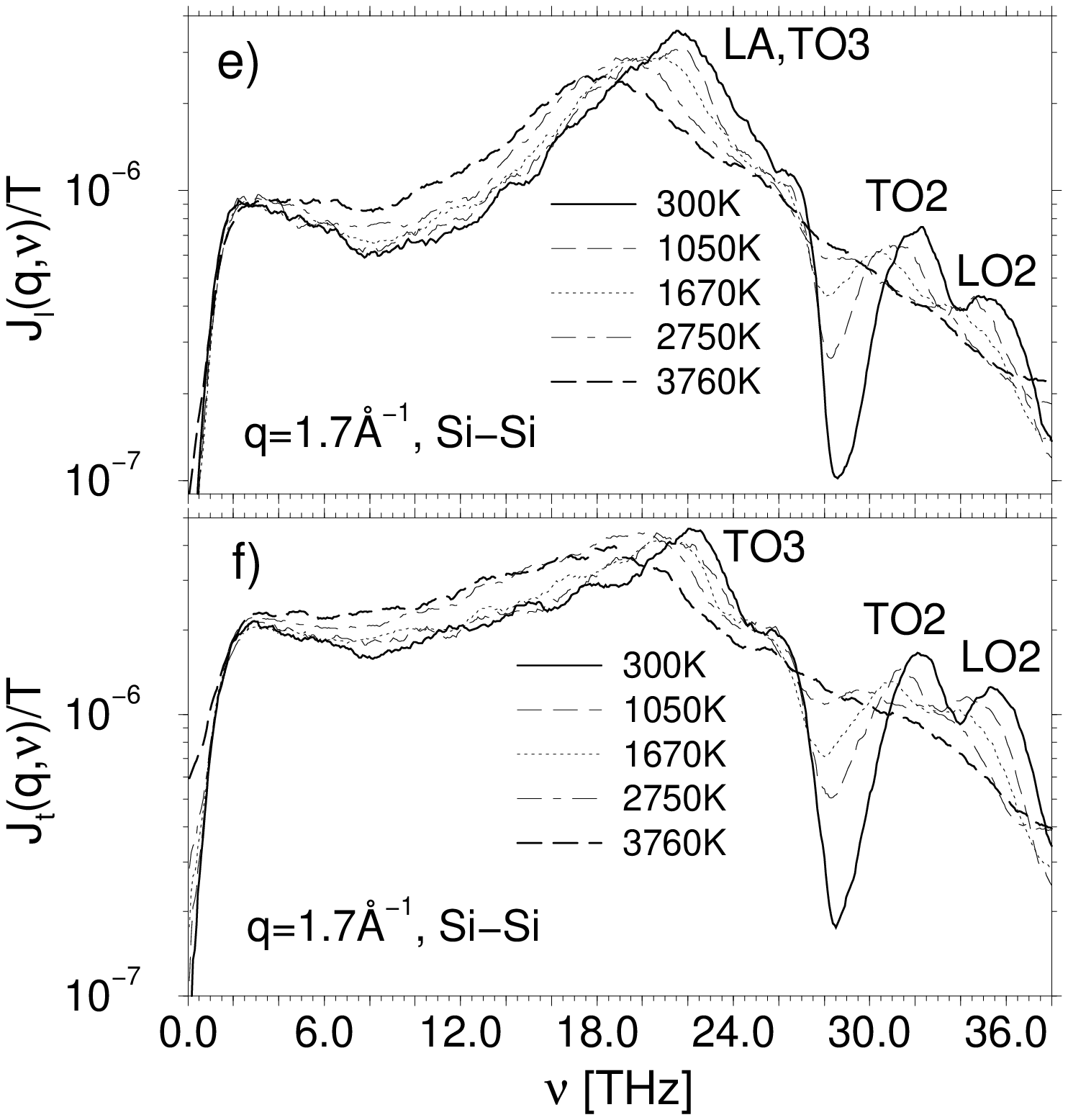,width=9cm,height=7.5cm}
\caption{The same as in Fig.~\protect\ref{fig6} but now for
         $q=1.7$~\AA$^{-1}$.}
\label{fig7}
\end{figure}

If the current correlation functions would behave as expected from the
harmonic approximation they would simply scale with temperature within
the classical treatment of our MD simulation. In order to see to what
extent a harmonic approximation holds, we have plotted in
Fig.~\ref{fig6} $J_l(q,\nu)/T$ and $J_t(q,\nu)/T$ at
$q=0.26$~\AA$^{-1}$ for the O--O, Si--O, and Si--Si correlations.  We
recognize from these figures that the different peaks corresponding to
the LA and TA excitations fall very well onto one master curve in the
whole temperature range $3760 \; {\rm K} \; \ge T \ge \; 300 \; {\rm
K}$.  Even for $T=3760$~K only weak deviations from the curve for
$T=300$~K are visible at low frequencies.  In contrast to these
acoustic excitations the optical band, i.e.~the excitations for $\nu >
20$~THz, exhibits strong deviations from a harmonic behavior. In the
following we denote the different excitations in the optical band as
LO1 and LO2 in the longitudinal case and as TO1, TO2, and TO3 in the
transverse case.  LO1 and LO2 correspond to the frequencies around
25.5~THz and 35.5~THz, respectively, and TO1, TO2, and TO3 correspond
to frequencies around $19$~THz, $22.3$~THz, and 32.2~THz, respectively.
(Note that we have been able to identify the different optical branches
as a function of $q$ by observing different peaks at approximately the
same frequency for different values of $q$.) It has been shown by
computer simulations \cite{tara97a,pasquarello98} and experiments
\cite{galeener79} that the LO2 and TO2 modes correspond to
intra--tetrahedral stretching modes, whereas the optical modes with
lower frequencies are mainly due to bending and rocking motions of
larger structural units.  With decreasing temperature the LO2 and TO2
peaks shift respectively from 32 to 36~THz and from 28.5 to 32~THz and
their amplitude increases. The mixing of the TO2 and LO2 modes becomes
evident in that a shoulder appears at $T=300$~K in $J_l(q,\nu)$ around
the frequency of the TO2 peak in $J_t(q,\nu)$ and also in $J_t(q,\nu)$
a feature is seen at the frequency of the LO2 mode. From the figure we
also recognize that the optical excitations at lower frequencies than
LO2 and TO2 have a weaker temperature dependence. This behavior is
reasonable since these modes are, in contrast to LO2 and TO2, more
extended since they originate from the inter--tetrahedral motion.
Nevertheless, also the LO1 peak increases in amplitude with decreasing
temperature and shifts to higher frequencies. The TO1 peak evolves into
a double peak structure (TO1 and TO3) in lowering the temperature.  In
Fig.~\ref{fig7} we show the current correlation functions scaled with
temperature at the higher wave--vector $q=1.7$~\AA$^{-1}$.  The current
correlation functions still scale approximately linear with temperature
for $\nu < 20$~THz. For these frequencies the strongest deviations from
such a behavior are found in the peak corresponding to the LA
excitations around $18$~THz.  Moreover the mixing of the LO2
excitations with the TO2 excitations is much stronger at
$q=1.7$~\AA$^{-1}$ than at $q=0.26$~\AA$^{-1}$.  Therefore, at $q$
values around $1.7$~\AA$^{-1}$ it makes no sense to distinguish these
excitations as transverse and longitudinal ones.

\begin{figure}[h]
\psfig{file=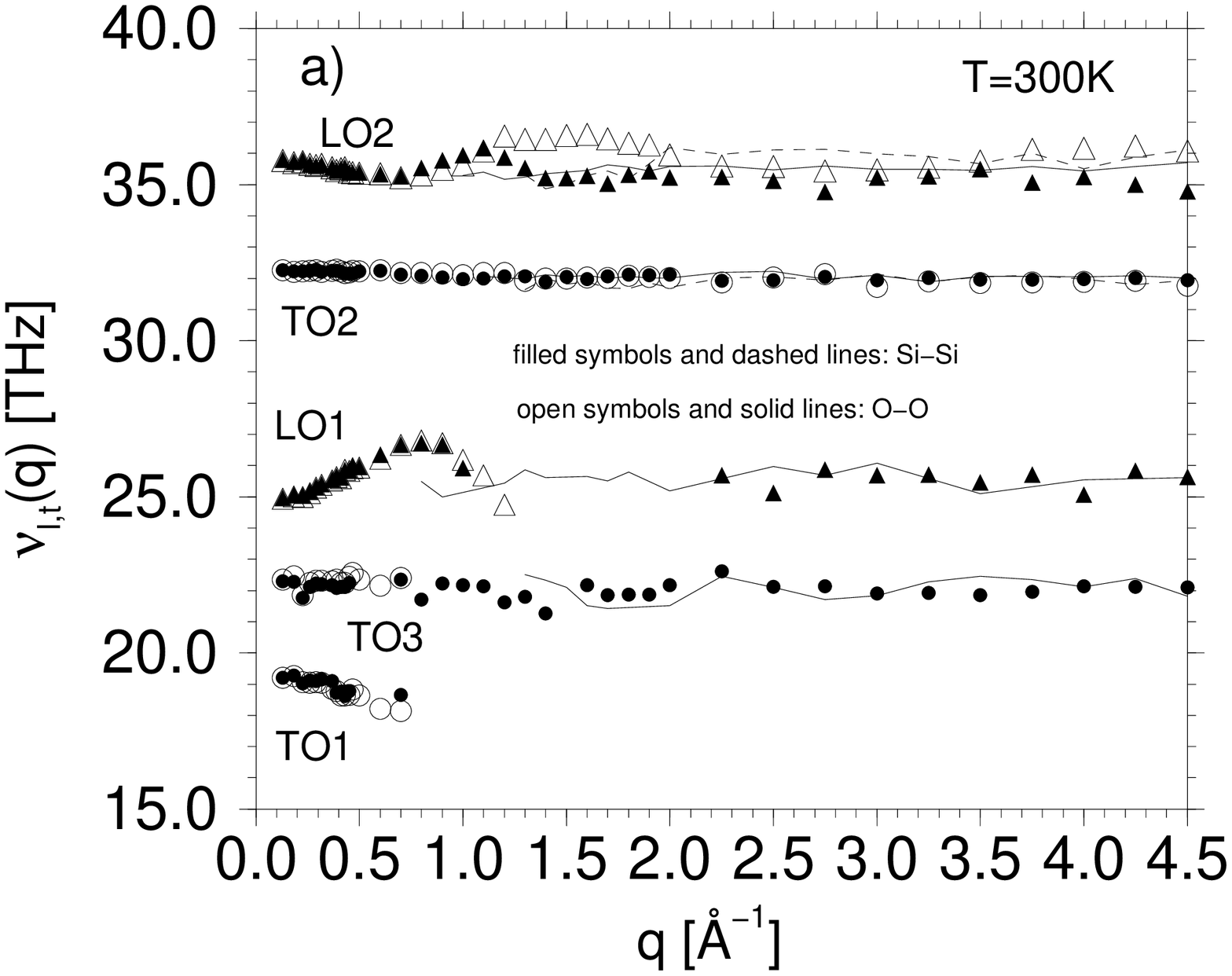,width=13cm,height=10.5cm}
\psfig{file=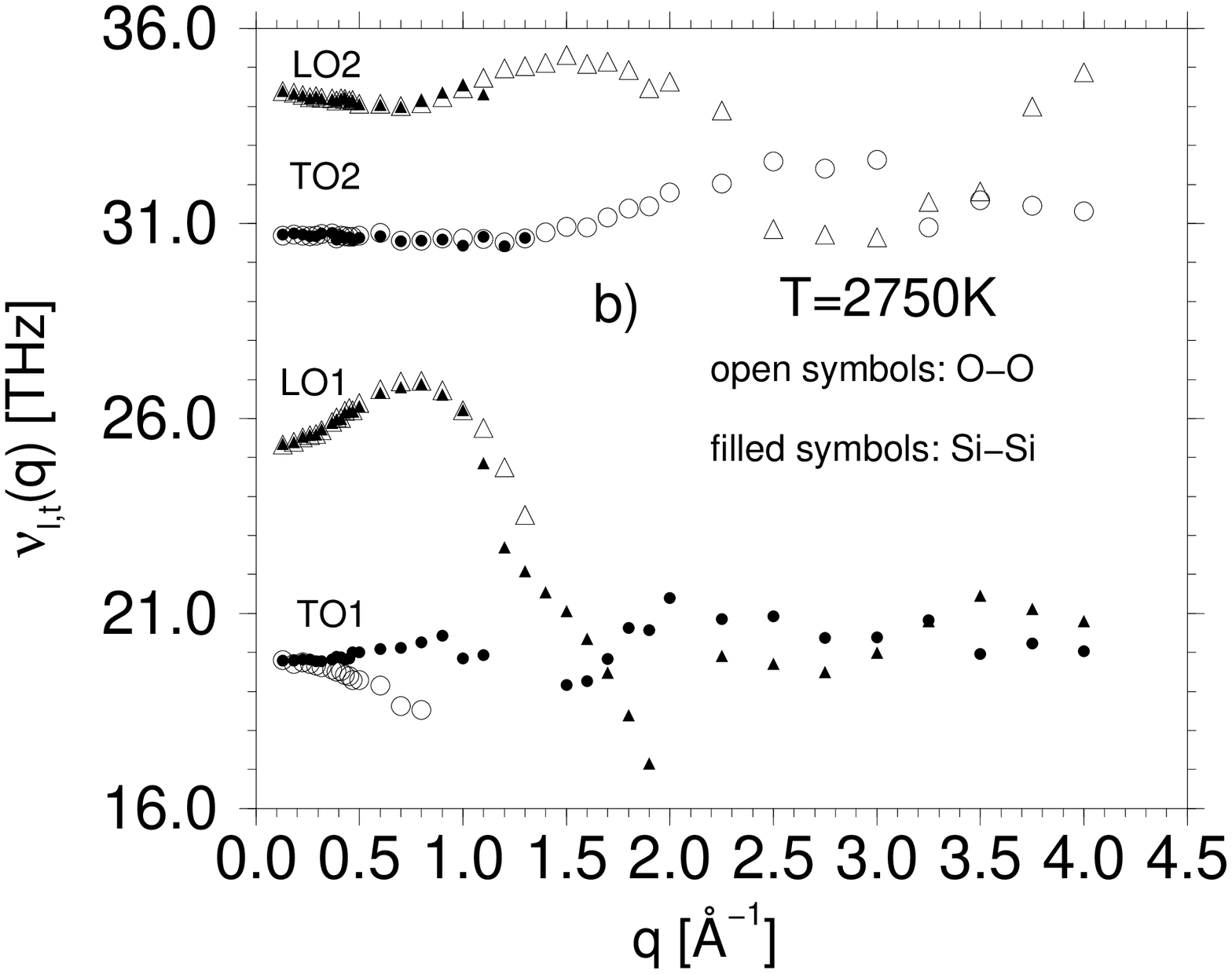,width=13cm,height=10.5cm}
\caption{Peak maximum position $\nu_{l,t}(q)$ of the excitations
        in the optical band for the Si--Si correlations 
        (filled symbols) and the O-O correlations (open symbols) 
        at a) $T=300$~K and b) $T=2750$~K. See text for the explanation of 
        the dashed and the solid line in a).}
\label{fig8}
\end{figure}

Figures~\ref{fig8} show the locations of all the peaks in
$J_{l,t}(q)$ that correspond to the optical excitations that can
be identified from the current spectra for the Si--Si and the
O--O correlations at the temperatures $T=300$~K and $T=2750$~K,
respectively. Whereas it is possible to distinguish at $T=300$~K the LO2
from the TO2 peaks for the whole $q$ range $0.13$~\AA$^{-1} \le q \le
4.5$~\AA$^{-1}$, this is not possible at $T=2750$~K for $q>2.0$~\AA$^{-1}$
because a strong mixing between LO2 and TO2 occurs for these values of
$q$.  The same is the case at $T=300$~K but at this temperature
the LO2 and TO2 branches can be distinguished from each other in that they
form a double peak stucture in the longitudinal and the transverse part.
In Fig.~\ref{fig8}a the mixing contributions at $T=300$~K are plotted
as the dashed and solid lines for the Si--Si and O--O correlations,
respectively: They show the occurrence of a longitudinal mode in the
transverse spectrum and, vice versa, the occurrence of a transverse mode
in the longitudinal spectrum. In this context the denotation of the modes
as longitudinal and transverse ones means that at very low values of $q$
they are expected to be purely longitudinal and transverse, respectively.

For $T=300$~K the optical branches at lower frequencies, i.e.~LO1, TO1,
and TO3, can be clearly identified in the Si--Si correlations and the
O--O correlations for $q<0.8$~\AA$^{-1}$.  At larger values of $q$ LO1
and TO3 appear as the dominant contribution in the Si--Si correlations,
whereas for the O--O correlations the spectra are dominated by the TO3
excitations in $J_l$ and the LO1 excitations in $J_t$.  At $T=2750$~K
the spectra are less pronounced, and we can identify only the TO1 and
the LO1 branch apart from the high frequency band above $30$~THz. They
are visible in the O--O and the Si--Si correlations for
$q<0.8$~\AA$^{-1}$. At higher $q$ only a broad flat band can be
observed in the O--O correlations in the frequency range $20 \; {\rm
THz} \; > \nu > \; 25 \; {\rm THz}$ and in the Si--Si correlations an
effective maximum appears around $\nu= 20.5$~THz.

\subsection{Density correlations}
\label{sec3.2}
Studying the density--density--correlation function in the 
$q$--$\nu$--domain, i.e.~the dynamic structure factor,
\begin{equation}
  S(q,\nu) = \frac{1}{N} \int_{-\infty}^{\infty} dt \; 
	     \left< \; \exp \left( i 2 \pi \nu t \right) \;
	     \sum_{k,l=1}^{N} 
	     \exp \left( i {\bf q} \cdot 
			 ({\bf r}_k (t) - {\bf r}_l (0))
                  \right) \right>,
             \label{sqnuedef}
\end{equation}
is of special interest because this quantity can be directly 
measured in neutron scattering experiments. $S(q,\nu)$ is 
related to the longitudinal current correlation function 
$J_l(q,\nu)$ by the simple equation
\begin{equation}
  S(q,\nu) = \frac{q^2}{4 \pi^2 \nu^2} \; J_l(q,\nu) \label{sqnue}
\end{equation}
which holds because of the continuity equation for particle number
conservation \cite{boon}. Equation (\ref{sqnue}) means that $S(q,\nu)$
and $J_l(q,\nu)$ contain the same information but features at lower
frequencies are strongly enhanced in $S(q,\nu)$ because of the factor
$1/\nu^2$. Moreover, $S(q,\nu)$ exhibits a quasielastic line around
$\nu=0$ whereas $J_l(q,\nu)$ approaches zero for $\nu \to 0$.
Therefore, one has to investigate density fluctuations in order to
understand the relationship between vibrational and relaxational dynamics,
which is one of the goals of the present sectioni, since the latter is
seen mainly at small $\nu$.

\begin{figure}[h]
\psfig{file=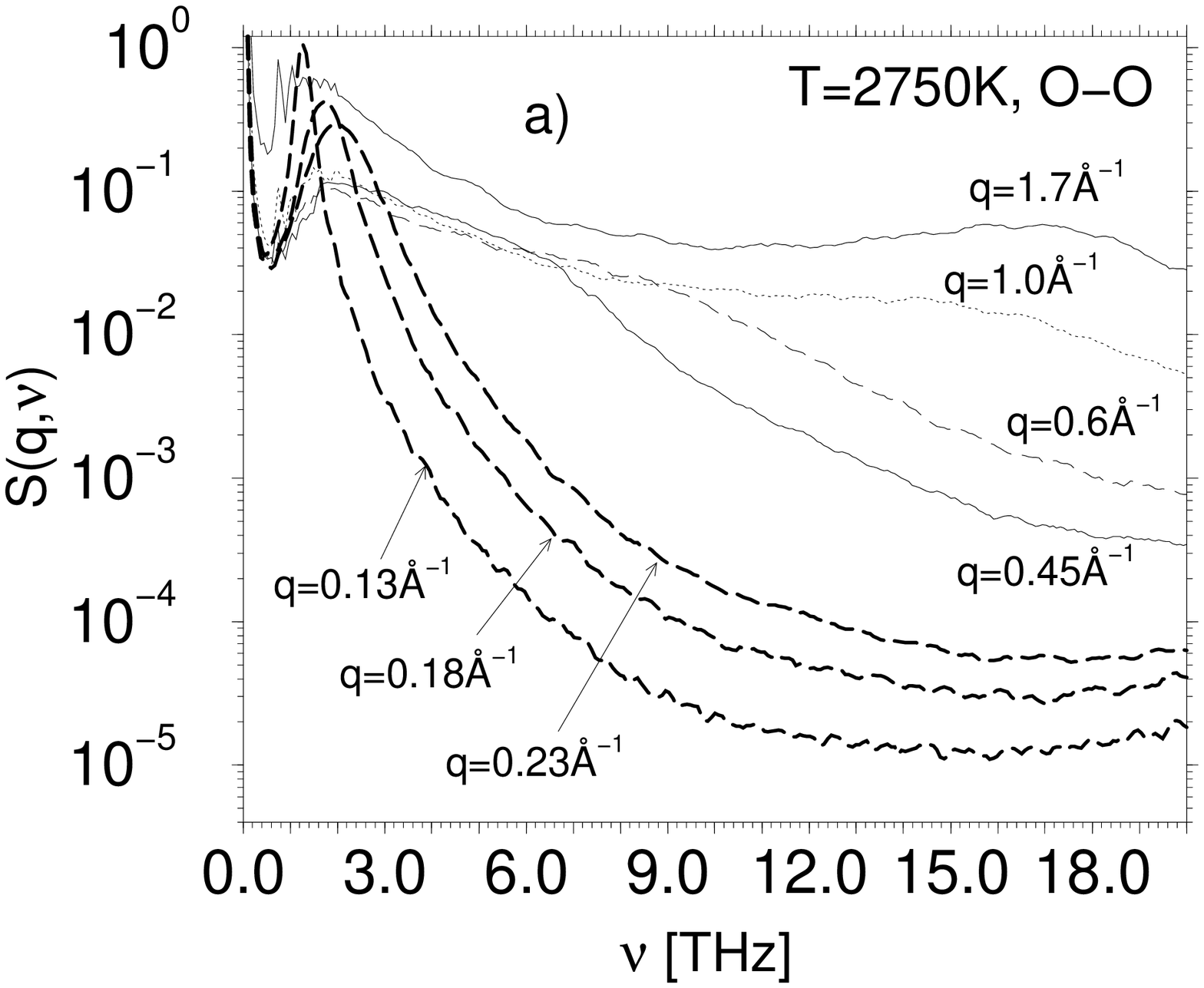,width=13cm,height=9.5cm}
\psfig{file=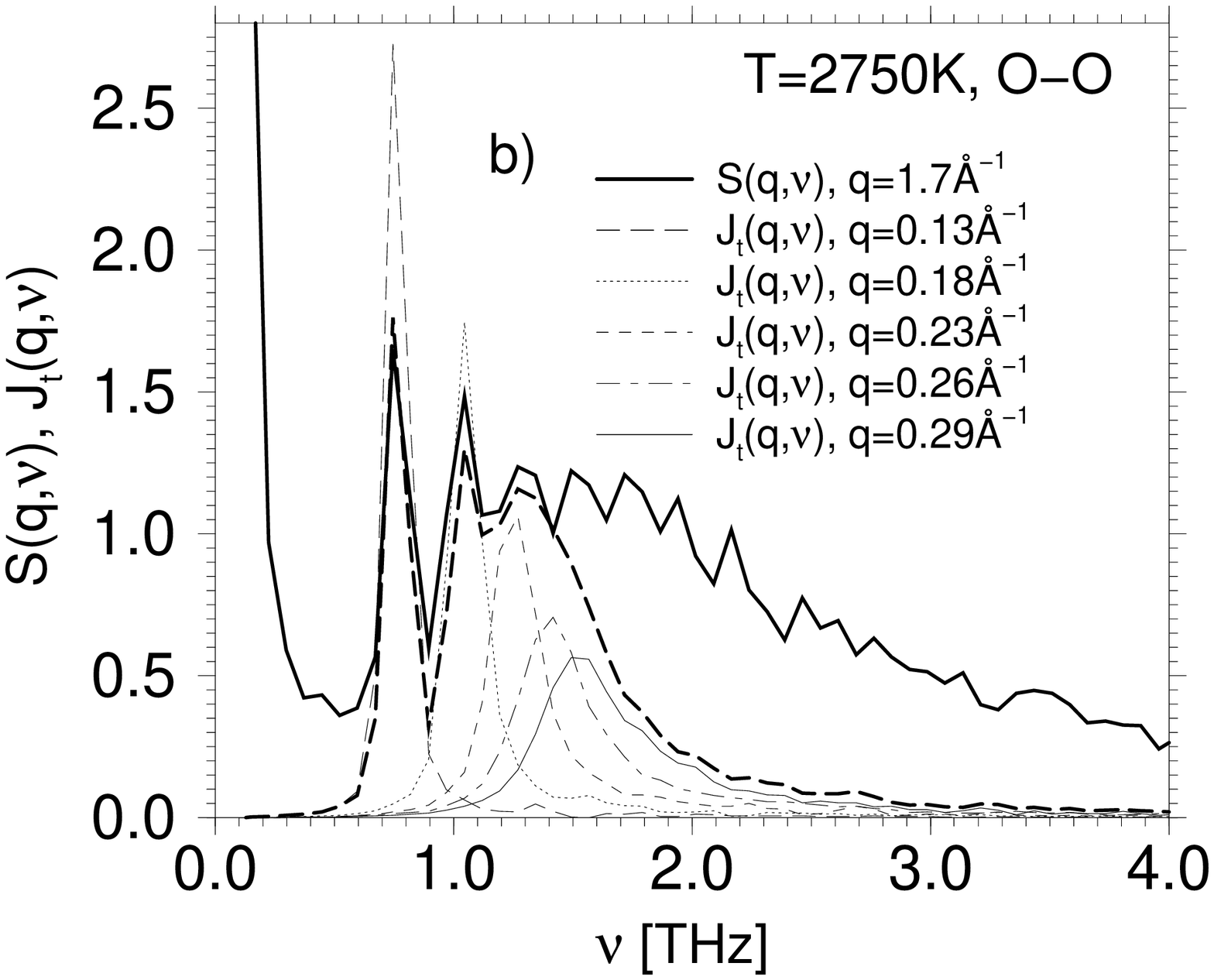,width=13cm,height=9.5cm}
\psfig{file=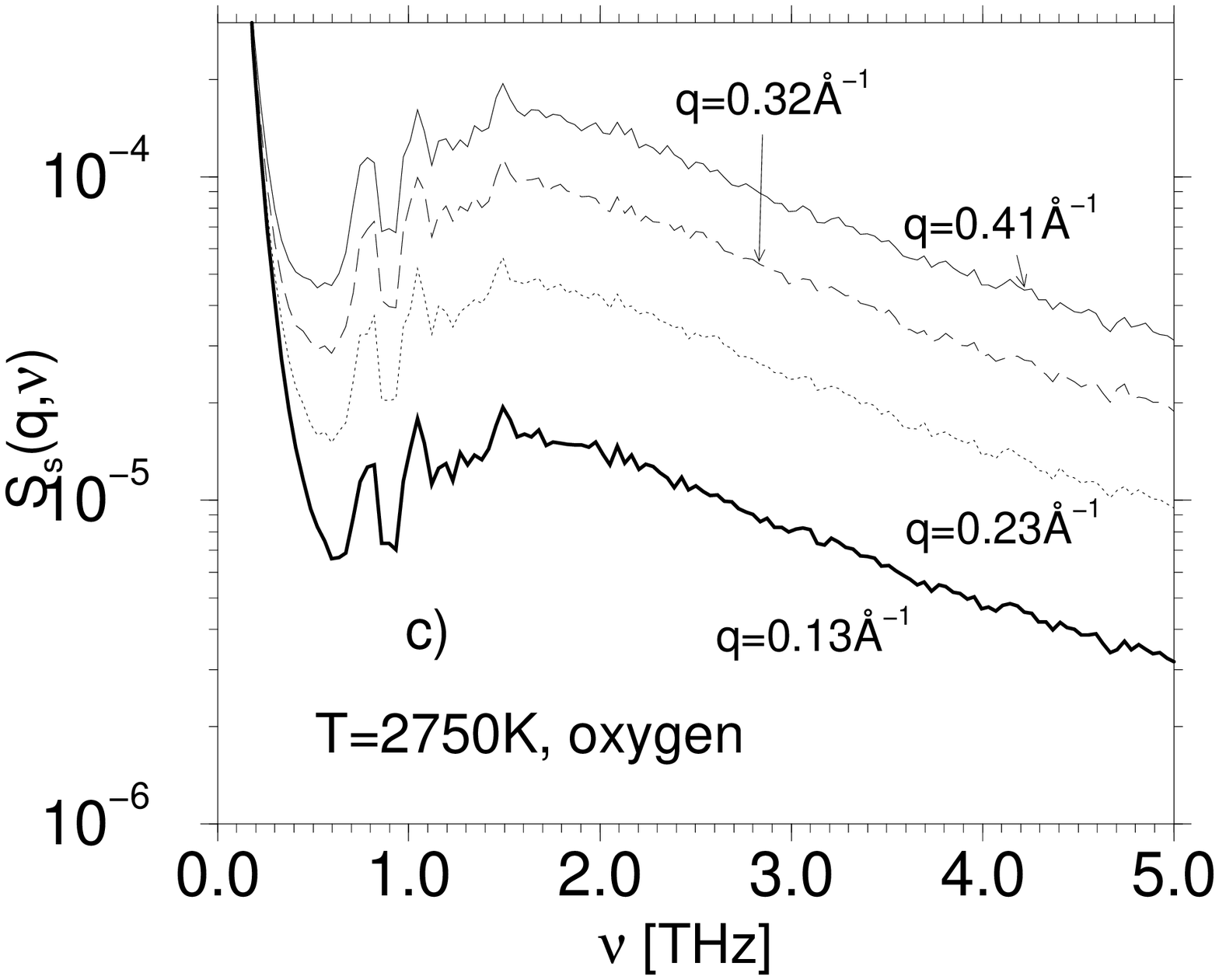,width=12cm,height=9.0cm}
\vspace*{-5mm}
\par
\caption{a) Dynamic structure factor $S(q,\nu)$ for the O--O 
	    correlations at $T=2750$~K for several values of $q$.
         b) $S(q,\nu)$ at $q=1.7$~\AA$^{-1}$ (bold solid line)
	    and $J_t(q,\nu)$
	    for the five lowest $q$ values of our simulation.
            The bold dashed line shows the sum of the latter
	    current correlation functions divided by 1.6.
         c) Self part of the dynamic structure factor 
	    $S_{{\rm s}}(q,\nu)$ for $q=0.13$~\AA$^{-1}$, 
	    $q=0.23$~\AA$^{-1}$, $q=0.32$~\AA$^{-1}$, and 
	    $q=0.41$~\AA$^{-1}$.}
\label{fig9}
\end{figure}

In Fig.~\ref{fig9}a $S(q,\nu)$ is shown for several values of $q$ at
$T=2750$~K. At the three lowest $q$ values of our simulation,
$q=0.13$~\AA$^{-1}$, $0.18$~\AA$^{-1}$ und $0.23$~\AA$^{-1}$, one sees
essentially one peak which corresponds to the longitudinal acoustic
excitations moving to higher frequencies with increasing $q$. At higher
values of $q$ the LA excitation is only visible as a shoulder until it
reaches $q=1.7$~\AA$^{-1}$ at which it can be identified as a broad
peak around $\nu= 17$~THz.  The reason why this excitation is
relatively hard to see is due to the fact that for $q>0.23$~\AA$^{-1}$
a second peak is present in $S(q,\nu)$ which is located nearly
independent of $q$ around $\nu=1.7$~THz.  This peak is the so--called
boson peak which is also seen experimentally for silica in Raman and
neutron scattering \cite{winterling75,foret96}.
From the figure we also see that to the left of the boson peak two
additional peaks are present. The location of these two peaks is at
$\nu=0.75$~THz and $\nu=1.05$~THz and we have checked that they are not
due to bad statistics nor due to artifacts of the Fourier
transformation. In order to discuss their origin we have plotted in
Fig.~\ref{fig9}b $S(q,\nu)$ at $q=1.7$~\AA$^{-1}$ and $J_t(q,\nu)$ for
the five lowest $q$ values of our simulation $q=0.13$~\AA$^{-1}$,
$0.18$~\AA$^{-1}$, $0.23$~\AA$^{-1}$, $0.26$~\AA$^{-1}$, and
$0.29$~\AA$^{-1}$. Also included in Fig.~\ref{fig9}b is the sum of
these transverse current correlation functions $J_{t,{\rm sum}}$ (bold
dashed line) which we have tried to scale onto $S(q,\nu)$ by dividing
it by $1.6$ in order to compare the shape of this function with that of
$S(q,\nu)$. From the comparison of $J_{t,{\rm sum}}$ with the dynamic
structure factor we can conclude that the main contribution to the low
frequency part of the boson peak comes from the coupling to the TA
modes at $q=0.13$~\AA$^{-1}$, $0.18$~\AA$^{-1}$, and $0.23$~\AA$^{-1}$.
The mechanism of how these modes couple to density fluctuations at
higher $q$, e.g.~at $1.7$~\AA$^{-1}$ in Fig.~\ref{fig9}b, is due to
elastic scattering since the energy of the scattered TA modes is
conserved.  We have seen before that the boson peak can only be
observed for $q > 0.23$~\AA$^{-1}$, which is exactly the region of $q$
in which the LA and TA peaks in $J(q,\nu)$ begin to overlap (see
Fig.~\ref{fig1}a).  That the
transverse part is of special importance is plausible since the
intensity of the TA peaks is a factor 6--8 higher than that of the LA
peaks in the current correlation functions at fixed $q$ (see
Fig.~\ref{fig1}). Note that around $\nu=1.0$~THz the band of acoustic modes is
not dense in our simulation because of the finite size of the
simulation box.  For this reason one would expect that the intensity
of the low frequency part of the boson peak is underestimated by our
simulation. But as it is demonstrated by Fig.~\ref{fig9}b this property
of the finite size system allows us to identify the influence of the
low $q$ TA modes on $S(q,\nu)$ at much larger $q$ and small $\nu$.

\begin{figure}[h]
\psfig{file=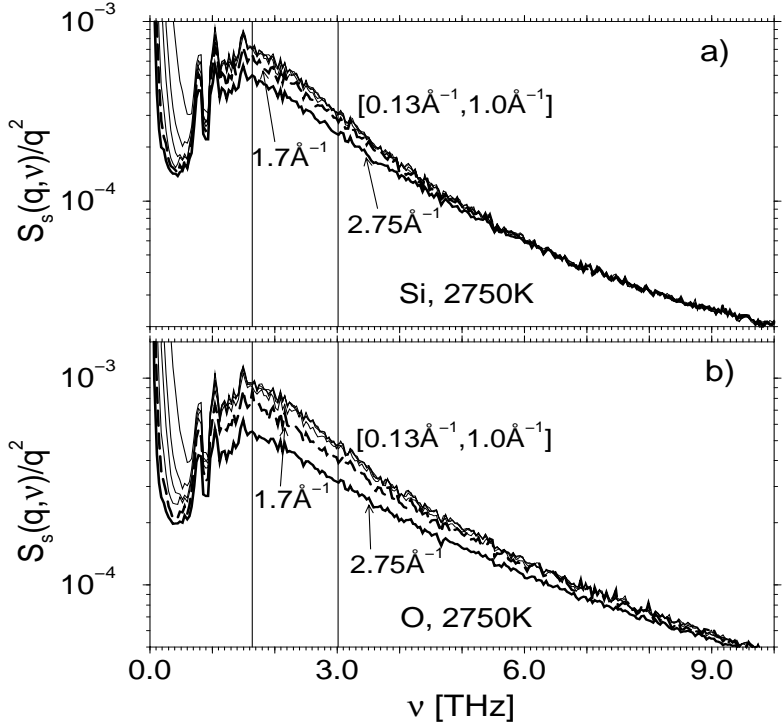,width=13cm,height=9.5cm}
\caption{Self part of the dynamic structure factor divided by $q^2$
	 for $q=0.13$~\AA$^{-1}$, $0.6$~\AA$^{-1}$, $1.0$~\AA$^{-1}$,
         $1.7$~\AA$^{-1}$, and $2.75$~\AA$^{-1}$ for a) Si and b) O.
         The vertical lines are at the frequencies $\nu=1.64$~THz and
         $3.02$~THz at which the $q$ dependence of $S(q,\nu)$ is shown
         in Fig.~\protect\ref{fignew11}.}
\label{fignew10}
\end{figure}

\begin{figure}[h]
\psfig{file=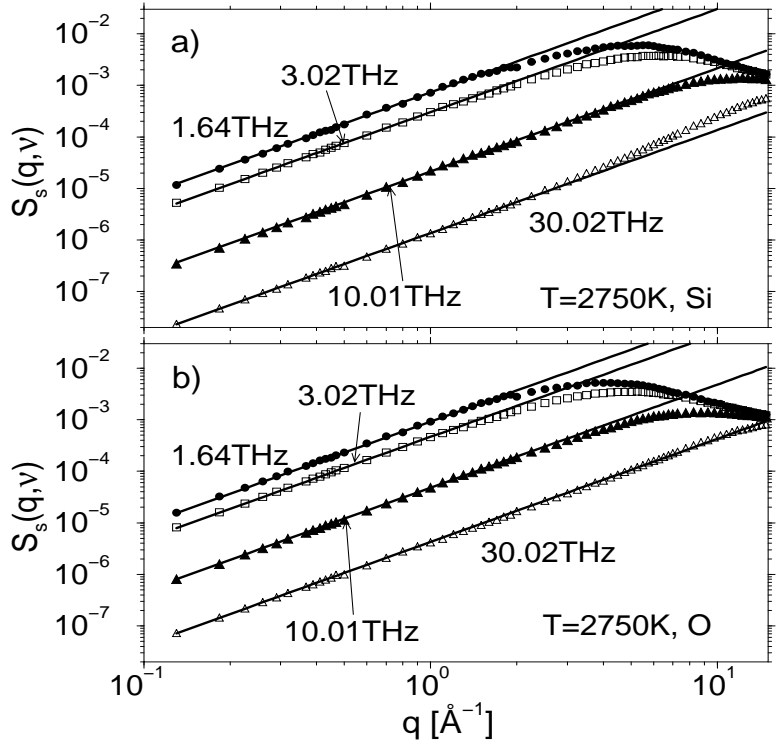,width=13cm,height=9.9cm}
\caption{$q$ dependence of $S_{{\rm s}}(q,\nu)$ for $\nu=1.64$~THz,
	 $3.02$~THz, $10.01$~THz, and $30.02$~THz, for a) Si,
	 and b) O. The bold straight lines are fits with $q^2$ laws.}
\label{fignew11}
\end{figure}

One might argue that the experimental result, that the boson peak is
observed in the Raman spectra of vitreous silica for $q$ values around
$10^{-3}$~\AA$^{-1}$, is in contradiction to our simulation in which we
find this peak in $S(q,\nu)$ only for $q>0.23$~\AA$^{-1}$.  But this is
probably due to the fact that in Raman scattering a strong coupling of
the light to the incoherent part of the density fluctuations is
present.
Indeed in our simulation the boson peak is also visible in the
self part of the dynamic structure factor $S_{{\rm s}}(q,\nu)$ which is
obtained from Eq.~(\ref{sqnuedef}) for $S(q,\nu)$ by taking into
account only the terms with $k=l$ in the sum. As can be seen in
Fig.~\ref{fig9}c even at $q=0.13$~\AA$^{-1}$, the smallest accessible
wave--vector, we observe the boson peak in $S_{{\rm s}}(q,\nu)$ and
this quantity also exhibits the sharp peaks around $\nu = 0.8$~THz and
$\nu=1.05$~THz that stem from the TA modes.  
Furthermore, the shape of the different curves in this
figure seems to be independent of $q$.  If this is the case this means
that $S_{{\rm s}}(q,\nu)$ can be factorized into a $q$ dependent part
and a purely frequency dependent part.  Indeed this has been predicted
recently in an analytic calculation by G\"otze and Mayr
\cite{goetze_draft} for a hard sphere system within the mode coupling
theory of the glass transition, and they found that the $q$ dependent
part is proportional to $q^2$.  In order to test whether this prediction holds
we have plotted $S_{{\rm s}}(q,\nu)/q^2$ for silicon and oxygen in
Figs.~\ref{fignew10}a and \ref{fignew10}b, respectively.  We recognize
that the curves for $0.13$~\AA$^{-1} \le q \le 1.0$~\AA$^{-1}$ fall
nicely onto one master curve in the whole frequency range $0.5$~THz$ \;
\le \nu \le 10$~THz whereas at larger $q$ small deviations from the
master curve are visible around the location of the boson peak,
$\nu=1.5$~THz.  To study the $q$ dependence of $S_{{\rm s}}(q,\nu)$ in
more detail we show in Fig.~\ref{fignew11} a double logarithmic plot of
this quantity at the frequencies $1.64$~THz, $3.02$~THz, $10.01$~THz,
and $30.02$~THz.  We see that fits with quadratic laws (bold lines in
the figures) hold very well at least for $q<2.0$~\AA$^{-1}$.  This
means that the whole spectrum scales with $q^2$ in this $q$ range. Note
that a similar behavior was also found in a simulation of ZnCl$_2$
\cite{foley95}.

\begin{figure}[h]
\psfig{file=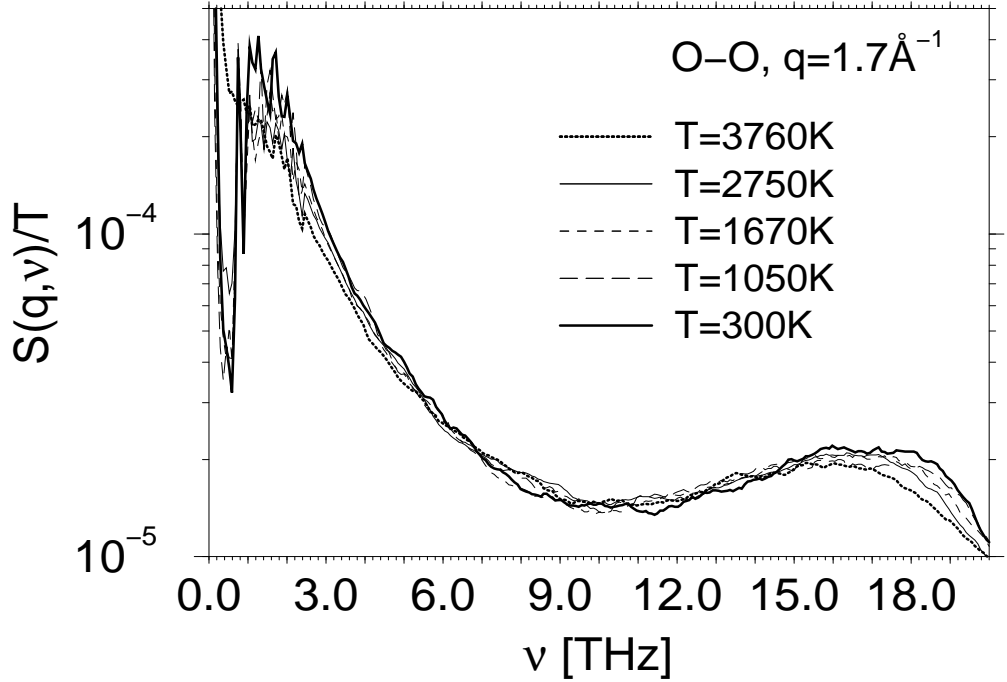,width=13cm,height=10.5cm}
\caption{$S(q,\nu)/T$ for the O--O correlations at $q=1.7$~\AA$^{-1}$
         for the different temperatures.}
\label{fig10}
\end{figure}

From the harmonic approximation one would expect that $S(q,\nu)$
scales with temperature. For this reason we have plotted in
Fig.~\ref{fig10} $S(q,\nu)/T$ as a function of frequency at
$q=1.7$~\AA$^{-1}$ for the temperatures $T = 3760$ K, $2750$ K, $1670$
K, $1050$ K, and $300$ K.  We recognize that the curves for the
different temperatures fall roughly on top of each other.  This means
that also in the region of the boson peak our silica model is quite
harmonic even at the relatively high temperature $T=3760$ K.  Note that
even at this temperature the boson peak feature is present in that a
shoulder is formed for frequencies above 0.5~THz.

\begin{figure}[h]
\psfig{file=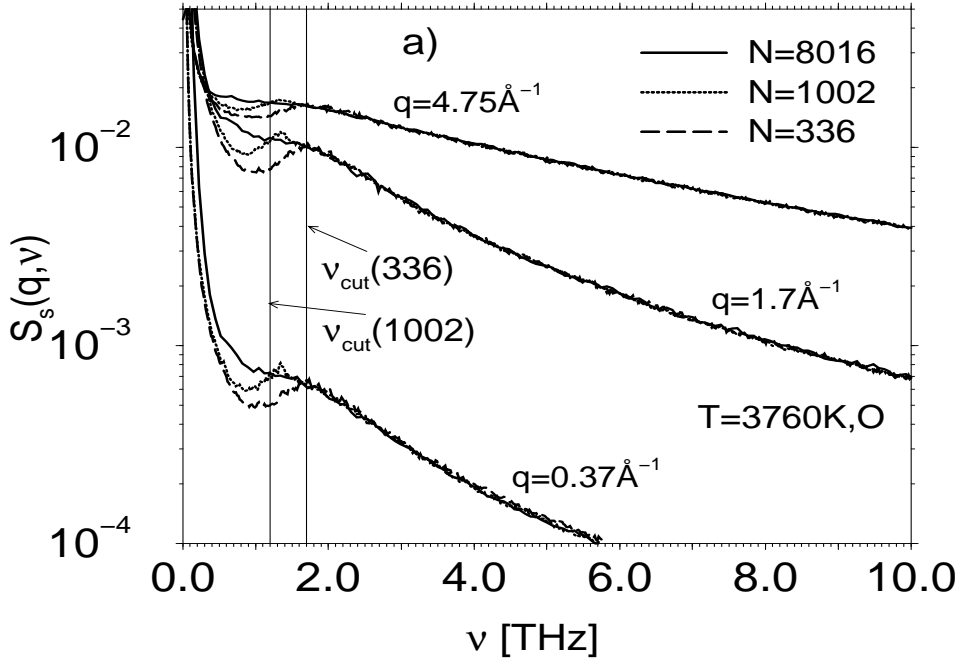,width=13cm,height=9.5cm}
\psfig{file=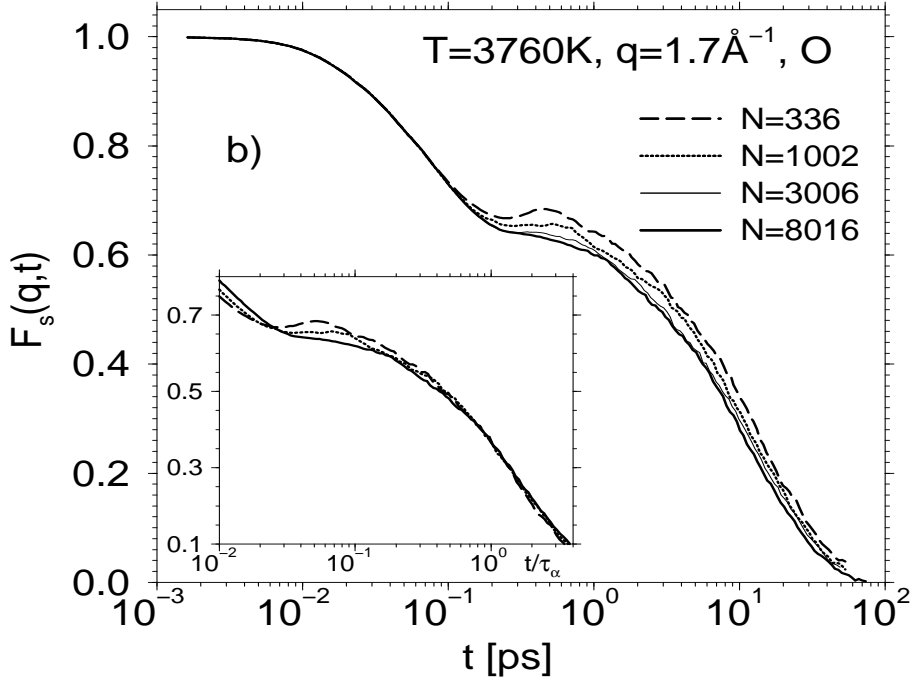,width=13cm,height=9.5cm}
\caption{a) The self part of the dynamic structure factor for oxygen 
	    for the different
            system sizes $N=8016, 1002, 336$ at the wave--vectors
            $q=0.37$~\AA$^{-1}$, $1.7$~\AA$^{-1}$, and
            $4.75$~\AA$^{-1}$, $T=3760$~K. See text for the 
            explanation of the vertical lines.
         b) Incoherent intermediate scattering function for oxygen
            for the system sizes $N=8016, 3006, 1002$, and $336$ at
            the wave--vector $q=1.7$~\AA$^{-1}$ and the temperature
            $T=3760$~K. The inset shows the same data (without the
            $N=3006$ curve), plotted versus the scaled time 
            $t/\tau_{\alpha}$.}
\label{fig11}
\end{figure}

If it is indeed true that TA modes with $q<0.2$~\AA$^{-1}$ couple to
density fluctuations at higher $q$ giving rise to certain features in
the boson peak, such as the additional peak at $\nu=0.8$~THz, then
these features should be absent if the system size is so small that it
does not allow the propagation of TA modes with $q<0.2$~\AA$^{-1}$. In
order to check this prediction we have calculated $S_{{\rm s}}(q,\nu)$
at $T=3760$~K for the system sizes $N=336$ and $N=1002$ in addition to
$N=8016$. As the same density as for $N=8016$ is used,
$\rho_m=2.37$~${\rm g}/{\rm cm}^3$, the sizes of the simulation boxes
are $L=16.80$~\AA~and $L=24.18$~\AA~for $N=336$ and $N=1002$,
respectively. Thus the smallest $q$ values are $q=0.37$~\AA$^{-1}$ and
$q=0.26$~\AA$^{-1}$. In Fig.~\ref{fig11}a we show the obtained $S_{{\rm
s}}(q,\nu)$ at $q=0.37$~\AA$^{-1}$, $1.7$~\AA$^{-1}$, and
$4.75$~\AA$^{-1}$ for the three system sizes.  Whereas the curves for
the different system sizes coincide for frequencies that are larger
than a weakly $N$ dependent frequency $\nu_{{\rm cut}}(N)$, for
$\nu<\nu_{{\rm cut}}(N)$ the magnitude of $S_{{\rm s}}(q,\nu)$
decreases with decreasing $N$.  Independent of $q$ we read off
$\nu_{{\rm cut}} \approx 1.7$~THz for $N=336$ and $\nu_{{\rm cut}}
\approx 1.2$~THz for $N=1002$. Both frequencies are marked by vertical
lines in Fig.~\ref{fig11}a.  $\nu_{{\rm cut}}(N)$ is indeed just
slightly below the frequency of the transverse excitation corresponding
to the lowest $q$ value determined by the size of the simulation box.
These frequencies are at $\nu=1.85$~THz for $N=336$ and at
$\nu=1.35$~THz for $N=1002$.  Thus this is evidence that the missing
of the TA modes with $q<0.2$~\AA$^{-1}$ causes the finite size effects
in the small systems. 

Due to the sum rule $\int d\nu \; S_{{\rm s}}(q,\nu) = 1$, the loss of
intensity in the boson peak below $\nu_{{\rm cut}}(N)$ has to be
``reshuffled'' to smaller frequencies leading to a broadening and an
increase of the intensity of the quasielastic line around $\nu=0$.
Since the quasielastic line is outside the frequency resolution of our
Fourier transformation the consequences in the change of the
quasielastic line can be observed better in the Fourier transform of
$S_{{\rm s}}(q,\nu)$, i.e.~the incoherent intermediate scattering
function $F_{{\rm s}}(q,t)$ which is defined as
\begin{equation}
  F_{{\rm s}}(q,t) = \frac{N_{\alpha}}{N} \int_{-\infty}^{\infty}
  d\nu \; \exp\left(2 \pi \nu t\right) S_{{\rm s}}(q,\nu)
  \quad \quad \quad \quad \alpha \in [{\rm Si, O}] \ .
  \label{ssqnueft}
\end{equation}
For oxygen this quantity is shown in Fig.~\ref{fig11}b at
$q=1.7$~\AA$^{-1}$ for different system sizes. (We mention that we have
also included a curve for $N=3006$ in the figure. The box length in
this case is $L=34.89$~\AA~corresponding to the lowest $q$ value
$q=0.18$~\AA$^{-1}$.)  We recognize from Fig.~\ref{fig11}b that with
decreasing system size the height of the plateau, which is the
Lamb--M\"ossbauer factor, increases and the $\alpha$--relaxation time
shifts to longer times.  Furthermore, the scattering functions for the
small systems show a pronounced oscillation for $t>0.2$~ps.  This can
be simply understood from the $\nu-$dependence of $S_{{\rm s}}(q,\nu)$: For
$N=8016$ this quantity shows a shoulder around $1.0$~THz which
corresponds to the monotonous decay of $F_{{\rm s}}(q,t)$. In the small
systems there is a peak in $S_{{\rm s}}(q,\nu)$ around $\nu_{{\rm cut}}(N)$
which corresponds to the oscillations in $F_{{\rm s}}(q,t)$ with a
period $1/\nu_{{\rm cut}}(N)$.  From the fact that the band of the
transverse acoustic modes is not dense for the region of small $q$ (see
Fig.~\ref{fig9}c) we expect also for $N=8016$ that the finite size
effects are present. But, since in Fig.~\ref{fig11}b the differences of
the curves for $N=3006$ and $N=8016$ are small, we can conclude that
the finite size effects play not an important role for $N=8016$, at
least for $T=3760$~K.  In order to investigate the influence of these finite size
effects on the $\alpha-$relaxation we define the $\alpha$--relaxation time
$\tau_{\alpha}$ as the time at which $F_{{\rm s}}(q,t)$ has decayed to
$1/{\rm e}$. $\tau_{\alpha}$ increases from $N=8016$ to $N=336$ by
about 40~\%. However, in the $\alpha$--relaxation regime the shape of
$F_{{\rm s}}(q,t)$ does not seem to depend on the system size, as can
be seen in the inset of Fig.~\ref{fig11}b where we have plotted
$F_{{\rm s}}(q,t)$ versus the scaled time $t/\tau_{\alpha}$. We see that in
the $\alpha$--relaxation regime the curves for the different system
sizes fall onto one master curve. This holds also for $F_{{\rm
s}}(q,t)$ for the silicon atoms.

\begin{figure}[h]
\psfig{file=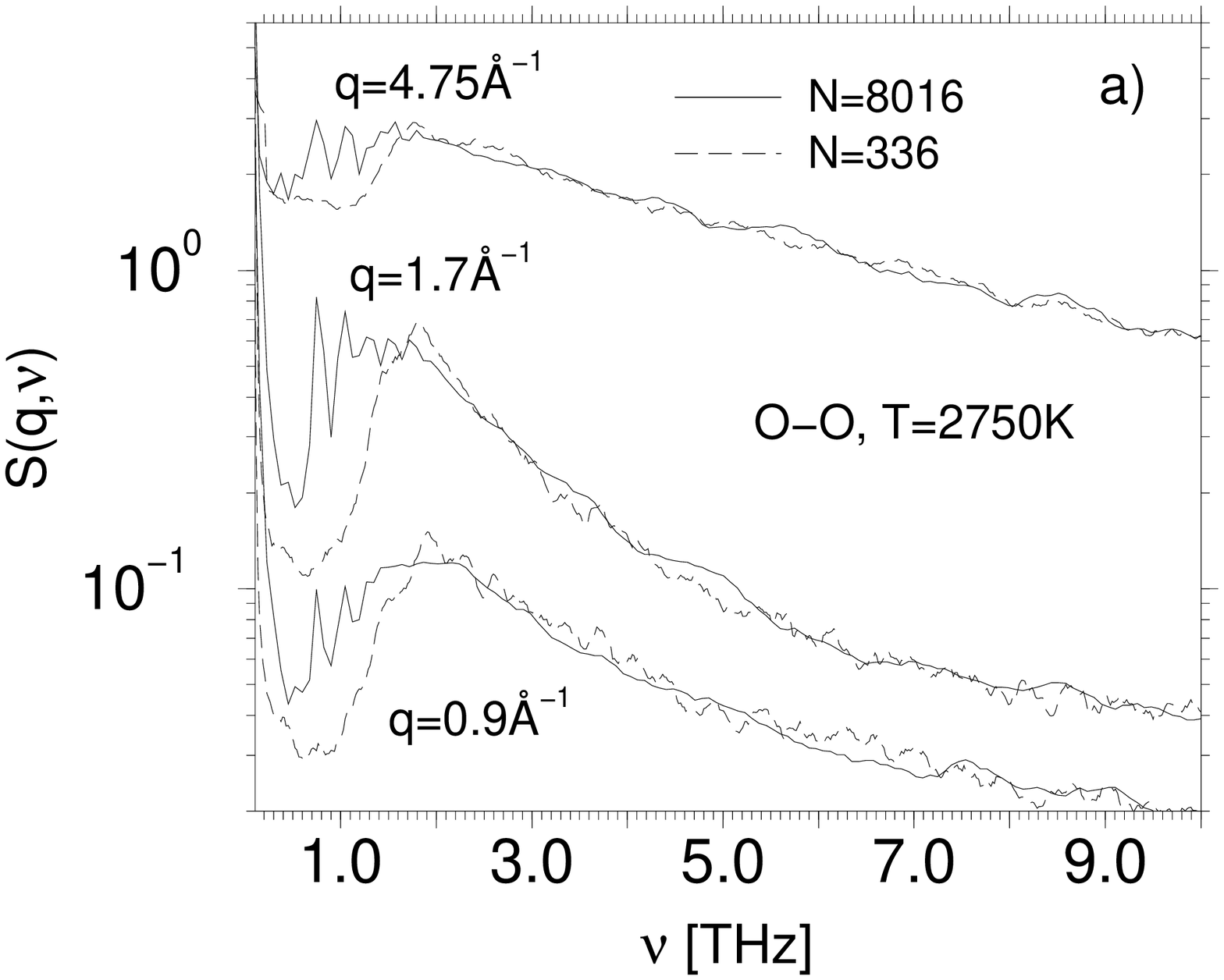,width=13cm,height=9.5cm}
\psfig{file=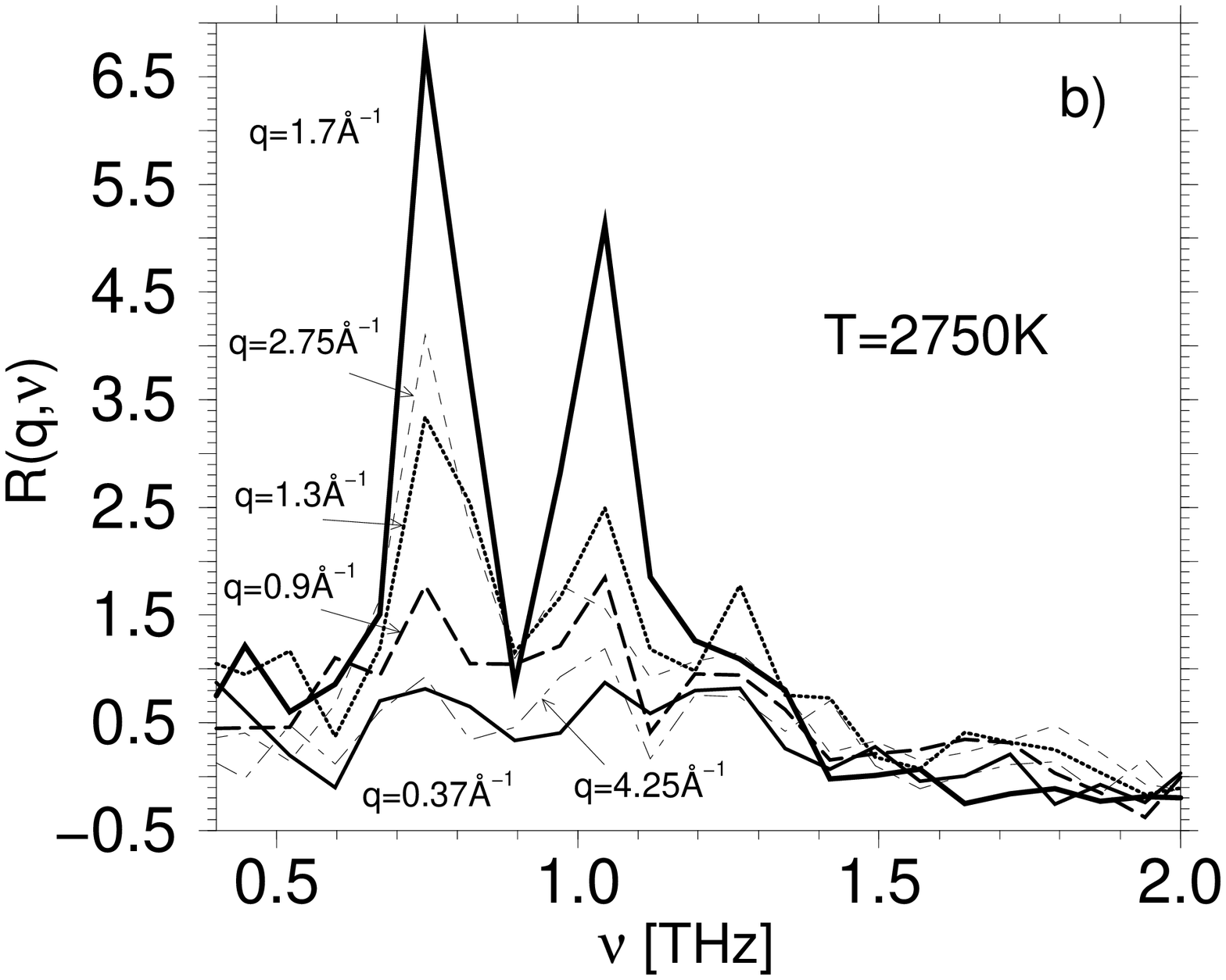,width=13cm,height=9.5cm}
\psfig{file=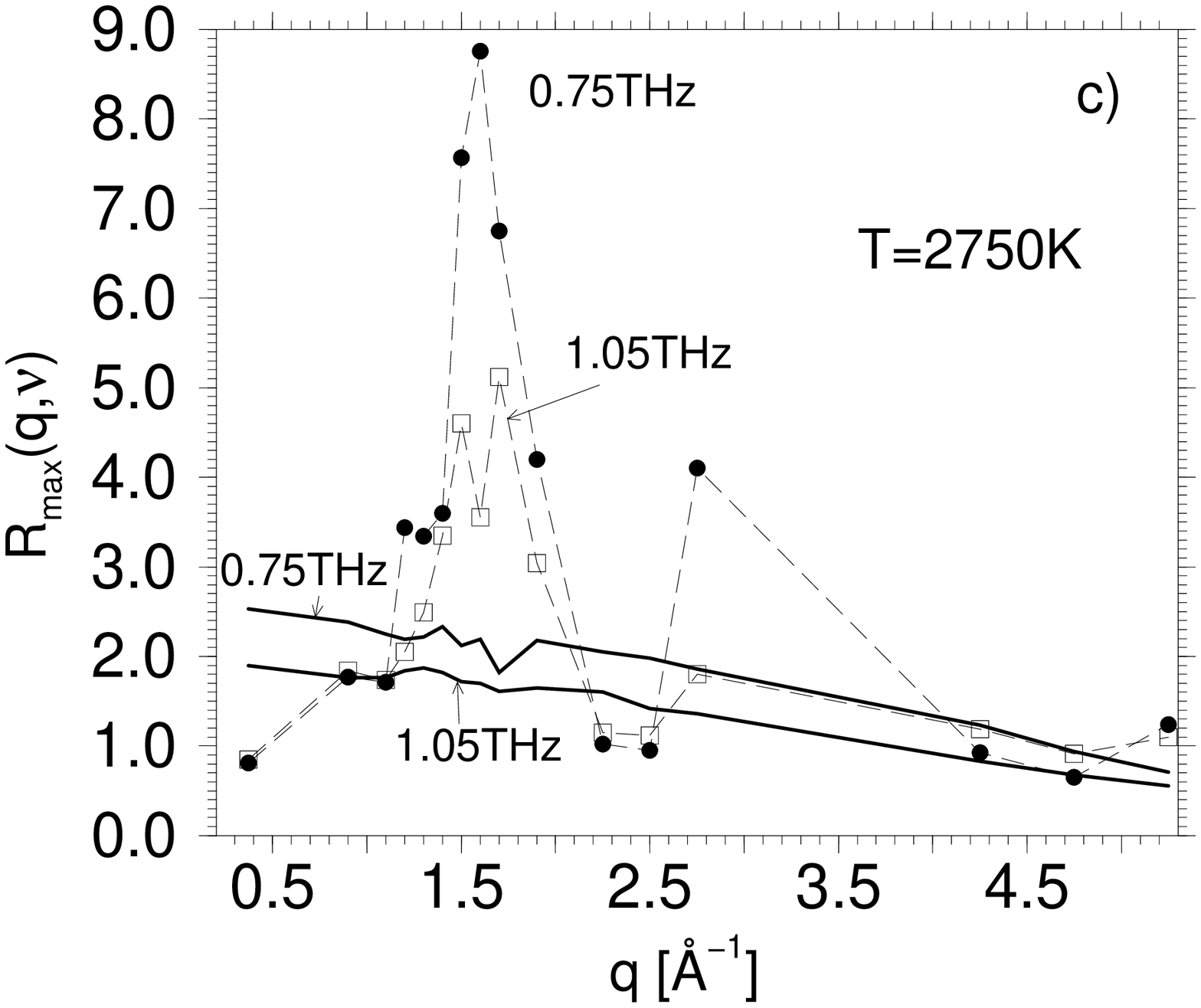,width=13cm,height=9.5cm}
\caption{a) The dynamic structure factor for the O--O correlations
	    for the system sizes $N=8016$ and $N=336$ at the
	    wave--vectors $q=0.9$~\AA$^{-1}$, $1.7$~\AA$^{-1}$, and
	    $4.75$~\AA$^{-1}$ and the temperature $T=2750$~K.
         b) $R(q,\nu)$ (definition see text) for several values of
	    $q$ as a function of frequency at $T=2750$~K.
	 c) $R(q,\nu)$ for $\nu=0.75$~THz (filled circles) and 
	    $\nu=1.05$~THz (open squares) as a function of $q$ at
	    $T=2750$~K. Also included is $R_{{\rm s}}(q,\nu)$ at 
	    the same frequencies obtained from $S_{{\rm s}}(q,\nu)$ 
	    (bold lines).}
\label{fig12}
\end{figure}

Of course, the finite size effects are also important in the total
dynamic structure factor. Figure~\ref{fig12}a shows $S(q,\nu)$ for the
two system sizes $N=8016$ and $N=336$ at $T=2750$~K and the three $q$
values $q=0.9$~\AA$^{-1}$, $1.7$~\AA$^{-1}$ and $4.75$~\AA$^{-1}$.  We
can again identify a cut--off frequency around $1.7$~THz below which
there is a loss of intensity in $S(q,\nu)$ for $N=336$. Note that the
sharp peaks at $\nu=0.75$~THz and $\nu=1.05$~THz are again not present
in the small system. Moreover, we recognize that the relative intensity
loss in the small system compared to the large system depends on
$q$. In order to quantify this $q$ dependence we define the ratio
\begin{equation}
  R(q,\nu) := \frac{S_{{\rm N=8016}}(q,\nu)}{S_{{\rm N=336}}(q,\nu)}
	    - 1 \label{rqnuofsqnu}
\end{equation} 
which is zero if the dynamic structure factor coincides for the two
system sizes. Figure \ref{fig12}b shows $R(q,\nu)$ for several values
of $q$. Its behavior underlines what we have said before that the low
frequency part of the boson peak is mainly due to the elastic
scattering of the two TA modes with the frequencies $\nu=0.75$~THz and
$\nu=1.05$~THz corresponding to the lowest $q$ values of our simulation
for the system size $N=8016$.  Obviously, the amplitudes of the peaks
in $R(q,\nu)$ do not change monotonously as a function of $q$. In order
to investigate this $q$ dependence $R(q,\nu)$ is plotted in
Fig.~\ref{fig12}c for the frequencies $\nu =0.75$~THz and
$\nu=1.05$~THz as a function of $q$.  $R(q,\nu)$ shows pronounced
maxima at $q=1.6$~\AA$^{-1}$, i.e.~in the vicinity of the location of
the first sharp diffraction peak in the static structure factor, and at
$q=2.8$~\AA$^{-1}$, which is the location of the second peak in
$S(q,\nu)$. Thus, the structural disorder on intermediate length
scales, i.e.~the length scale introduced by two connected
SiO$_4$--tetrahedra, is most relevant for the scattering of the TA
modes with $q<0.2$~\AA$^{-1}$.  Also included is $R_{{\rm s}}(q,\nu)$
for the two frequencies which is obtained by putting in $S_{{\rm
s}}(q,\nu)$ instead of $S(q,\nu)$ into the definition
(\ref{rqnuofsqnu}).  The incoherent part $R_{{\rm s}}(q,\nu)$ decreases
monotonously with $q$ which is plausible since the finite size effects
should vanish at very large values of $q$ corresponding to small length
scales.

\section{Summary and Conclusions}
\label{sec4}

We have done molecular dynamics simulations in order to investigate the
high frequency dynamics of amorphous silica. The results which we have
presented in this paper have been for the fully equilibrated liquid and
the glass state. The frequency range we have studied is $0.5$~THz~$ <
\nu < 40$~THz for wave--vectors with magnitude $q \ge 0.13$~\AA$^{-1}$
(limited by the size of the simulation box). 

In a first step we have discussed the properties of the longitudinal
and transverse current correlation functions.  At low frequencies we
have identified propagating longitudinal acoustic (LA) and transverse
acoustic (TA) modes the maxima of which move to higher frequencies with
increasing $q$. The amplitude of the TA peaks is a factor 6--8 larger
than that of the LA peaks at a fixed value of $q$ which is a first
indication for the importance of the transverse dynamics in silica.
Whereas the LA peak is separated quite well from the TA peak on the
frequency axis at $q=0.13$~\AA$^{-1}$, both peaks begin to overlap at
higher $q$. The $q$ region at which the LA and TA peaks begin to
overlap significantly can be seen as a crossover region from a
regime where the longitudinal and transverse modes exhibit only a weak
interaction to a regime where a strong interaction between different
modes is present.  One important sign of this is that 
the qualitative shape of the spectra starts to
change gradually around $q=0.6$~\AA$^{-1}$:  The LA peaks are still 
well--defined, but they are
now sit on top of a flat background.  The acoustic band in
$J_t(q,\nu)$ shows a similar behavior in that it evolves into a broad
plateau from about 3 to 11~THz.  The observation that the acoustic
modes are located on top of a flat background for intermediate values
of $q$ has recently been found by G\"otze and Mayr \cite{goetze_draft}
as an essential result in an analytic calculation of the spectra of
hard sphere systems in their glass state within the mode--coupling
theory of the glass transition.  Within their theory G\"otze and Mayr
have explained the existence of the flat background spectrum by the
presence of inelastic phonon scattering where a mode decays into two
modes due to anharmonicity. In the same sense we can understand the
flat background spectrum in silica as the manifestation of multiphonon
excitations.

By reading off the peak maxima in $J_l(q,\nu)$ and $J_t(q,\nu)$ as a
function of $q$ one obtains dispersion like functions $\nu_l(q)$ and
$\nu_t(q)$ for the longitudinal and transverse part, respectively.
$\nu_l(q)$ shows an approximately linear behavior for wave--vectors up
to $0.3$~\AA$^{-1}$.  On the other hand the full width at half maximum
$\Gamma_l(q)$ of the LA peaks is well decribed by a quadratic law for
$q\le 0.5$~\AA$^{-1}$.  This means that to a good approximation the
system behaves like an isotropic elastic medium up to
$q\approx0.3$~\AA$^{-1}$ with respect to the longitudinal sound modes.
Furthermore, $\nu_l(q)$ exhibits a quasi Brillouin zone at $q_{{\rm
m}}/2$ where $q_{{\rm m}}$ is the location of the second sharp
diffraction peak in the static structure factor correponding to length
scales of intratetrahedral distances. Also $\nu_t(q)$ shows
approximately a linear behavior at small $q$. But the data for
$\Gamma_t(q)$ cannot be described by a quadratic law. Instead,
$\Gamma_t(q)$ is fitted well with a $q^{2.5}$ law from which we
conclude that the TA excitations are stronger damped than expected from
an isotropic elastic medium.  For $q>0.8$~\AA$^{-1}$ $\nu_t(q)$ becomes
flat, and in the same range the TA excitations become strongly
overdamped in that they reach a Ioffe--Regel limit, i.e.~$\Gamma_t(q)$
is of the order of $\nu_t(q)$. We emphasize that there are no
pecularities in the qualitative behavior of $\nu_l(q)$ and $\nu_t(q)$
for our silica model since these functions look very similar for simple
liquids \cite{boon}.

From the two lowest $q$ values of our simulation, $q=0.13$~\AA$^{-1}$
and $0.18$~\AA$^{-1}$, we have determined the apparent high frequency sound
velocities for the different temperatures and find that they reproduce
the light scattering data by Vo--Tanh {\it et al.}~\cite{votanh} very
well. Thus this is another example that the BKS model is able to
reproduce reliably the experimental data of amorphous silica.

Most of the results we have summarized up 
to now for the acoustic band are for the silica melt at $T=2750$~K. But
we have seen that in the temperature range $300$~K $\; \le T \le \;
3760$~K the spectra scale to a good approximation linearly with
temperature for frequencies $\nu<20$~THz, a behavior which is expected
if the harmonic approximation is valid. The linear scaling is not valid
for the complex optical bands above 20~THz. For these excitations the
spectrum becomes more pronounced with decreasing temperature. If one
plots the current correlation functions divided by temperature one
recognizes that the peak maxima corresponding to the
intratetrahedral stretching modes LO2 and TO2 move to higher
frequencies and also their amplitude increases with decreasing
temperature. This is just the result of the fact that the particles are
more localized the lower the temperature is, which causes
anharmonicities to dissapear.  Nevertheless, due to the disorder,
especially for the LO2 and TO2 modes, a strong mixing occurs between the
longitudinal and the transverse part for $q>0.3$~\AA$^{-1}$ which can
be clearly identified at $T=300$~K. So, these wave--vectors are no
longer good quantum numbers, and if one treats disordered structures
within the harmonic approximation one has to keep in mind that this
approximation only describes the ``mixing contributions'' in an 
effective way.

In a second step we have discussed density fluctuations by means of the
dynamic structure factor $S(q,\nu)$. For $q>0.23$~\AA$^{-1}$ this
quantity exhibits a boson peak which is located nearly $q$ independent
around $\nu_{{\rm BP}}=1.7$~THz at $T=2750$~K. The boson peak
excitations coexist with the LA modes since the latter is visible also
at frequencies above $\nu_{{\rm BP}}$. Since the boson peak has a much
larger intensity than the LA peak, e.g.~a factor of 7--8 for
$q=1.0$~\AA$^{-1}$, the LA excitations are visible only as a shoulder
in $S(q,\nu)$.  Only if the LA peak has moved to high frequencies,
e.g.~to $17$~THz at $q=1.7$~\AA$^{-1}$, one observes this peak as a
second peak in addition to the boson peak in the dynamic structure
factor.  In the low frequency part of the boson peak two sharp peaks
are present at $\nu=0.75$~THz and $\nu=1.05$~THz which are due to the
elastic scattering of the TA modes with $q=0.13$~\AA$^{-1}$ and
$q=0.18$~\AA$^{-1}$, respectively.  We will discuss them in more detail
below. Also the dynamic structure factor $S(q,\nu)$ scales for
frequencies around $\nu_{{\rm BP}}$ roughly with temperature in the
range $3760$~K$ \; \ge \; T \ge 300$~K.  Note that at $T=3760$~K the
boson peak feature can be only seen as a shoulder which grows out of
the quasielastic line. The fact that the boson peak can be seen even
at such high temperatures supports the view of Ref.~\cite{goetze_draft}
that this feature becomes visible as soon as temperature is around $T_c$,
which for our system is around 3330~K~\cite{horbach99a}.

For wave--vectors up to about $1$--$2$~\AA$^{-1}$ the self part of the
dynamic structure factor $S_{{\rm s}}(q,\nu)$ exhibits a factorization
into a frequency dependent and wave--vector dependent part whereby the
latter is proportional to $q^2$. Apart from the fact that this property
of $S_{{\rm s}}(q,\nu)$ has also been found in a MD simulation of
ZnCl$_2$ \cite{foley95} it is remarkable that this result comes out
of the mode coupling calculation of Goetze and Mayr
\cite{goetze_draft}.  So this is another important feature which is
reproduced by this theoretical approach.

In order to get more insight into the boson peak feature we have done
simulations also for smaller system sizes than the normally used system
with $N=8016$~particles. We have found strong finite size effects in
the low frequency part of the boson peak which can be characterized by
a frequency $\nu_{{\rm cut}}(N)$ below which there is a lack of
intensity in the dynamic structure factor. The frequency $\nu_{{\rm
cut}}(N)$ decreases with increasing system size $N$ and is essentially
independent of $q$.  The reason for these finite size effects is due to
the absence of the TA excitations with $q<0.2$~\AA$^{-1}$ in the small
systems since the smallest $q$ value of our simulations with $N=1002$
and $N=336$ particles are $0.26$~\AA$^{-1}$ and $0.37$~\AA$^{-1}$,
respectively.  In the time domain, i.e.~in the incoherent intermediate
scattering function $F_{{\rm s}}(q,t)$, the finite size effects cause
an increase of the Lamb--M\"ossbauer factor and of the $\alpha-$relaxation 
time. This is a consequence of the sum rule $\int d\nu \;
S_{{\rm s}}(q,\nu) = 1$ since the missing of intensity for
$\nu<\nu_{{\rm cut}}(N)$ has to be ``reshuffled'' to smaller
frequencies.  Because of the abrupt decrease of $S_{{\rm s}}(q,\nu)$
below $\nu_{{\rm cut}}(N)$ for small $N$ we observe quite pronounced oscillations for
$t>0.2$~ps in $F_{{\rm s}}(q,t)$ whereby the period of these
oscillations is approximately equal to $\nu_{{\rm cut}}(N)$. Note that 
a similar
behavior was also found in a MD simulation by Lewis and Wahnstr\"om
\cite{lewis94} for a model of orthoterphenyl in which the interactions
between the molecules are described by a Lennard--Jones potential.
These authors have suggested that a disturbance that propagates through
the system will leave and reenter the box due to the periodic boundary
conditions after a time $L/c$, where $L$ is the size of the box and $c$
is the typical velocity of the sound wave. This mechanism then produces
an echo, i.e.~an additional signal which produces the slowed down decay
of the correlation functions like $F_{{\rm s}}(q,t)$. We have also
suggested this explanation recently for silica \cite{horbach96}, but we
think now that this explanation for the finite size effects is not the
correct one.  Instead, the general argument is that in small enough
systems (with a smallest wave--vector with magnitude $q_{{\rm s}}$)
parts of the vibrational spectrum are missing below a frequency
$\nu_{{\rm cut}}(N)$ because of the coupling to wave--vectors with
$q<q_{{\rm s}}$ occuring in an infinite system. In a Lennard--Jones
system such a coupling is reflected in the flat background spectrum
which was shown to be present by Mazzacurati {\it et
al.}~\cite{mazzacurati96}.  In the case of silica there is in addition
the coupling which arises from the elastic scattering of transverse TA
modes with small $q$ by the structural disorder.

One may speculate that the strength of the boson peak
in silica is due the strong coupling of the TA exciations to the
longitudinal part. The stiffness of the tetrahedral SiO$_2$ network
introduces strong restoring forces which allow the propagation of 
shear waves with a large amplitude even at relatively high temperatures.
We have shown that the strongest scattering of the TA modes is at
$q\approx1.6$~\AA$^{-1}$. This is in agreement with suggestions in
the literature that the boson peak is caused by the interactions
of sound modes with coupled rotations of several 
tetrahedra \cite{buchenau86,tara97b}. A possible mechanism of
this interaction would be that the coupled rotations of the
tetrahedra enable the change of the polarization of transverse 
acoustic modes, so that they contribute to the density fluctuation
spectrum, i.e.~constituting at least part of the boson peak.

Acknowledgments: We thank A. Latz, M. Letz, W. G\"otze, G. Ruocco,
and F. Sciortino for many stimulating discussions during this work.
We also thank D. Vo--Tanh for providing us with the light scattering
data of the sound velocities.  This work was supported by BMBF Project
03~N~8008~C and by SFB 262/D1 of the Deutsche Forschungsgemeinschaft.
We also thank the HLRZ J\"ulich for a generous grant of computer time
on the CRAY-T3E.

\begin{references}
\bibitem{peak_max}
In order to obtain the peak maxima $\nu_{\alpha}(q)$ we have used the
formula
%
\begin{equation}
  J_{\alpha}(q,\nu) =
  \frac{A_{\alpha}(q) \nu^2}{(\nu^2 - \nu_{\alpha}^2)^2 +
        (B_{\alpha}(q) \nu)^2} .  \label{jdho}
\end{equation}
%
which contains the fit parameters $A_{\alpha}(q)$, $B_{\alpha}(q)$, and
$\nu_{\alpha}(q)$. Note that Eq.~(\ref{jdho}) has the same form as the
solution of the linearized Navier--Stokes equation for $J_l(q)$, when
the heat mode is neglected \cite{boon}. We emphasize that we have used
Eq.~(\ref{jdho}) only as a fit formula to determine $\nu_{\alpha}(q)$.
Therefore, we have applied Eq.~(\ref{jdho}) only in a small interval
around the location of the maximum position $\nu_{\alpha}(q)$.
\end{references}
\end{document}